\documentclass{article}
\usepackage[affil-it]{authblk}
\usepackage[english]{babel}

\usepackage[letterpaper,top=2cm,bottom=2cm,left=3cm,right=3cm,marginparwidth=1.75cm]{geometry}

\usepackage{amsmath}
\usepackage{graphicx}
\usepackage{color,xcolor}

\usepackage[colorlinks=true, allcolors=blue]{hyperref}
\usepackage{float}

\title{From policy to prediction: Forecasting COVID-19 dynamics under imperfect vaccination
}
\author[1,2]{Xiunan Wang}
\author[1]{Hao Wang\footnote{The corresponding author. Email: hao8@ualberta.ca}}
\author[3]{Pouria Ramazi} 
\author[1,4]{Kyeongah Nah} 
\author[1,5]{Mark Lewis}

\affil[1]{Department of Mathematical and Statistical Sciences, University of Alberta, Edmonton, AB T6G 2G1, Canada}
\affil[2]{Department of Mathematics, University of Tennessee at Chattanooga, Chattanooga, TN 37403, USA
}
\affil[3]{Department of Mathematics and Statistics, Brock University, St. Catharines, ON L2S 3A1, Canada}
\affil[4]{National Institute for Mathematical Sciences, Daejeon 34047, Korea}
\affil[5]{Department of Biological Sciences, University of Alberta, Edmonton, AB T6G 2G1, Canada}

\begin{document}
\maketitle

\begin{abstract}
\noindent
Understanding the joint impact of vaccination and non-pharmaceutical interventions on COVID-19 development is important for making public health decisions that control the pandemic.
Recently, we created a method in forecasting the daily number of confirmed cases of infectious diseases by combining a mechanistic ordinary differential equation (ODE) model for infectious classes and a generalized boosting machine learning model (GBM) for predicting how public health policies and mobility data affect the transmission rate in the ODE model \cite{Wang2021}.
In this paper, we extend the method to the post-vaccination period, accordingly obtain a retrospective forecast of COVID-19 daily confirmed cases in the US, and identify the relative influence of the policies used as the predictor variables. 
In particular, our ODE model contains both partially and fully vaccinated
compartments and accounts for the breakthrough cases, that is, vaccinated individuals can still get infected.
Our results indicate that the inclusion of data on non-pharmaceutical interventions can significantly improve the accuracy of the predictions. 
With the use of policy data, the model predicts the number of daily infected cases up to 35 days in the future, with an average mean absolute percentage error of 34\%, which is further improved to 21\% if combined with human mobility data. 
Moreover, similar to the pre-vaccination study, the most influential predictor variable remains the policy of restrictions on gatherings. 
The modeling approach used in this work can help policymakers design control measures as variant strains threaten public health in the future.
\end{abstract}

\section{Introduction}\label{Introduction}
Since COVID-19 was characterized as a pandemic by World Health Organization (WHO) on March 11, 2020, it has spread to 224 countries and territories. The United States is the country most affected by COVID-19, with 20,629,998 confirmed cases and 369,897 deaths by the end of December 2020 \cite{worldometer}. 
Mass vaccination against COVID-19 started on December 20, 2020 in the US. 
As of Dec 13, 2021, $72.6\%$ of the US population have received at least one dose of vaccine, $60.9\%$ have been fully vaccinated, and $16.5\%$ have been given a booster shot \cite{owidcoronavirus}. 
Except another small peak in April, the weekly number of new cases kept decreasing nearly monotonically, since mid January 2021 until June 2021 \cite{owidcoronavirus}, which brought a faint hope that the COVID-19 pandemic might be brought under control soon, although this hope has been dashed by the emergence of new COVID-19 variants. 
Indeed, even when vaccines are available, control of COVID-19 indispensably relies on some non-pharmaceutical interventions (NPIs), such as testing, contact tracing, facial coverings, protection of elderly people, school closing, workplace closing, cancellation of public events, restrictions on gatherings, public transport closing, stay at home requirements \cite{owidcoronavirus}.
Thus it is necessary and urgent to understand the joint impact of vaccination and NPIs on COVID-19 spread in order to provide guidance for policymakers to control the pandemic.

Transmission dynamics is a useful tool to serve this purpose, as it can assess both the direct and indirect impact of vaccinations on the disease spread  \cite{eichner2017direct, halloran1991direct}. 
Dynamical models have been used in studying COVID-19 vaccine prioritization, hypothetical vaccination strategies or the resource allocation, such as the intensity of NPIs needed to balance with a restricted number of vaccines available \cite{bubar2021model, brett2020transmission, buckner2021dynamic, saad2020immune, macintyre2021modelling, matrajt2021vaccine, han2021time, li2020modeling}. 
In terms of the future projection, the majority of studies provide only qualitative insights rather than quantitative estimates. 
Quantitative forecasting of the future transmission in the post-vaccination era can be realized only if we predict the number of COVID-19 confirmed cases based on vaccination and NPI policy data.

In our recent pre-vaccination modeling work \cite{Wang2021}, we employed a hypothesis-free machine-
learning algorithm to estimate the transmission rate based on NPI data, and in turn forecast the daily number of confirmed cases in the US for the pre-vaccination period using a mechanistic disease model. 
We also investigated the impact of different types of policy and mobility data on the predictions and found that restrictions on gatherings is the most influential variable \cite{Wang2021}. 
In this paper, we use the same method as in \cite{Wang2021} to make a retrospective forecast of the daily number of confirmed cases in the US for the post-vaccination period and investigate the joint impact of vaccination and NPIs. 
More specifically, we build a hybrid model consisting of a mechanistic ordinary differential equation (ODE) and a generalized boosting machine learning model (GBM). 
The ODE model contains two vaccinated compartments: partially vaccinated and fully vaccinated, and it accounts for the case that vaccinated individuals can still get infected. 
Then the impact of vaccination is reflected implicitly when the ODE model gives simulation results. 
The NPIs serve as predictor variables of the GBM to predict the transmission rate. 
Before we run GBM to make predictions, we use the inverse method that we created in \cite{Wang2021} to obtain a time series of daily transmission rates which are fed into GBM as the response variable. 
The GBM is trained based on these predictor and response variable data and produces predictions of future transmission rates given future NPIs.
Using the predicted transmission rates from the GBM, the ODE model gives the predicted number of daily confirmed cases.
In \cite{Wang2021}, we have shown that including NPI policy data can greatly improve the accuracy of the predictions. 
We were curious to see whether this is also the case for the post-vaccination period. 
To this end, in addition to the scenario where only policies are used as the predictors of the GBM, we have considered two other scenarios where human mobility is also used.
This forecasting approach capturing the joint impact of vaccination and NPIs can hopefully be applied to other countries that are suffering terrible situations caused by SARS-CoV-2 and its variant strains as well as other infectious diseases. 

The remaining paper is organized as follows. 
In Section \ref{Methods}, we present data collection, model formulation, parameter estimation, and prediction methodology.
In Section \ref{Results}, we present the results. 
In Section \ref{Discussion}, we provide concluding remarks and propose possible future work.

\section{Methods}\label{Methods}
\subsection{Data collection}
In this study, we collected daily data from April 4, 2020 to April 5, 2021 that cover part of both pre-vaccination and post-vaccination periods in the US. We obtained the number of confirmed cases of COVID-19, the number of partially vaccinated and fully vaccinated individuals in the US and policy indices in each state of the US from the website of {\it Our World in Data} \cite{owidcoronavirus} (\url{https://ourworldindata.org/coronavirus}), the six categories of human mobility data in the US from the official website of Google \cite{Google} (\url{https://www.google.com/covid19/mobility/}), and deaths, recovered and active cases in the US from the worldometer website \cite{worldometer} (\url{https://www.worldometers.info/coronavirus/country/us/}). 

\begin{figure}[hp]
\centering
\includegraphics[width=0.49\textwidth]{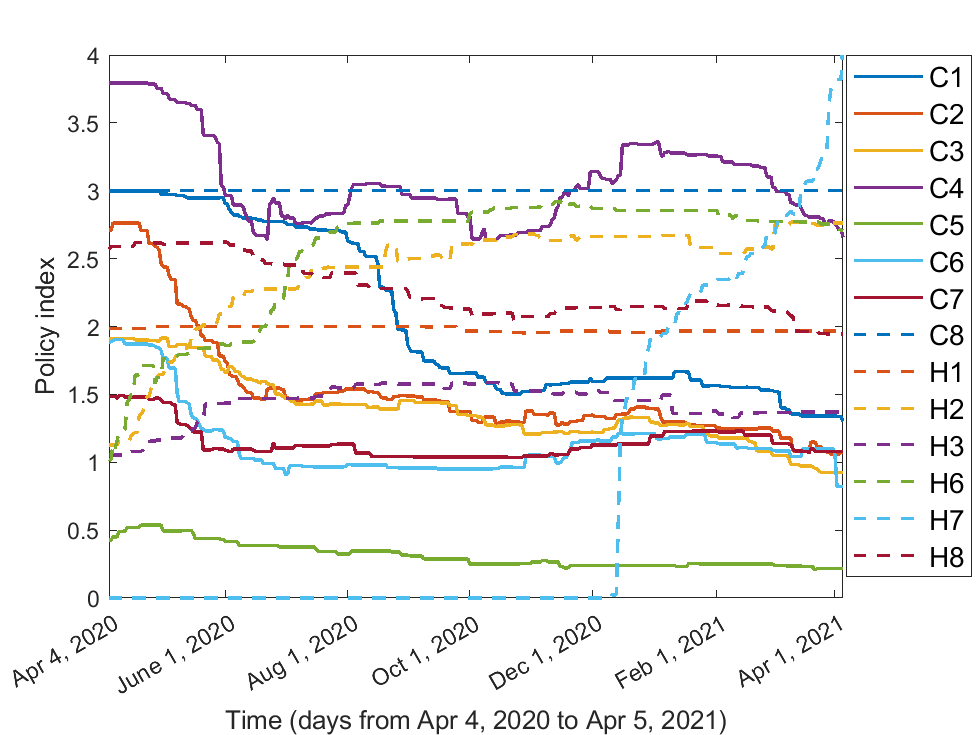}
\includegraphics[width=0.49\textwidth]{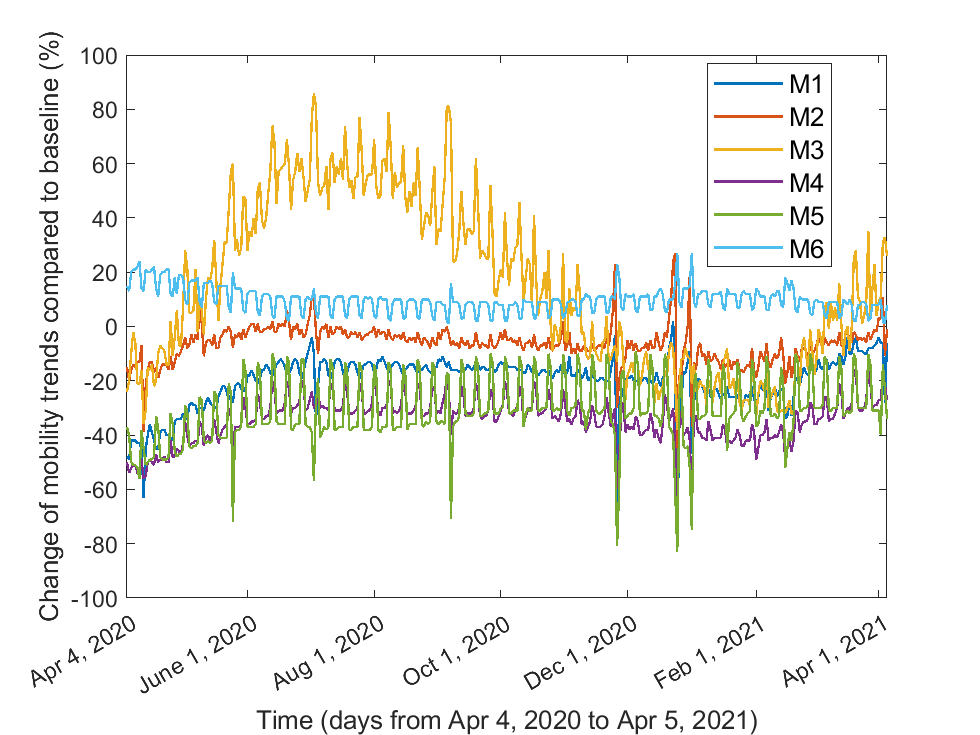}
\caption{\label{PolicyFig} Policy and mobility data in the US from Apr 4, 2020 to Apr 5, 2021.}
\end{figure}

We derived the time-series index data for containment policies (beginning with ``C") and health policies (beginning with ``H") in the US by taking an average of the corresponding policy indices over all the 50 US states as well as Washington D.C., weighted by their populations.
The containment policies included school closing (denoted by C1), workplace closing (C2), cancel public events (C3), restrictions on gatherings (C4), close public transport (C5), stay at home requirements (C6), restrictions on internal movement (C7), and international travel controls (C8). The health policies included public information campaigns (H1), testing policies (H2), contact tracing (H3), facial coverings (H6), vaccination delivery policy (H7), and protection of elderly people (H8). The labels H4 and H5 represent emergency investment in healthcare and investment in vaccines, respectively, which are not available \cite{owidcoronavirus}. Note that in addition to those policy data in our pre-vaccination work \cite{Wang2021}, here we take into account the vaccination delivery policy (H7) as well since we focus on the post-vaccination case in the current paper.
Human mobility data included changes of mobility trends ($\%$), compared to the baseline level 0, in retail and recreation (M1), grocery and pharmacy (M2), parks (M3), transit stations (M4), workplaces (M5), and residential (M6). The time series of these 14 policy indices and human mobility data are shown in Figure \ref{PolicyFig}.

\subsection{Model formulation}
Our model extends the SEIAR (Susceptibles, Exposed, symptomatic Infected, Asymptomatic infected, Removed) model in \cite{Wang2021} by incorporating two new compartments: the individuals who are partially vaccinated (denoted by $V_1$), and the individuals who are fully vaccinated ($V_2$). The susceptible individuals getting their first dose of vaccines will enter the $V_1$ compartment, and the individuals in the $V_1$ compartment will enter the $V_2$ compartment if they are fully vaccinated. 
According to CDC statistics, a small part of vaccinated individuals can still get infected, which is the so-called “breakthrough cases” (\url{https://www.cdc.gov/vaccines/covid-19/health-departments/breakthrough-cases.html}). We describe such breakthrough cases by incorporating incidence terms in the equations of $V_1$ and $V_2$, with relative risks of infection being $\epsilon_1$ and $\epsilon_2$, respectively, for partially vaccinated and fully vaccinated individuals, due to imperfect vaccination protection. Therefore, the susceptible individuals ($S$), the partially vaccinated individuals ($V_1$), and the fully vaccinated individuals ($V_2$) will all enter the exposed compartment (E) if they are infected by symptomatic infected individuals (I), asymptomatic infected individuals (A), or the exposed individuals (E). The transmission rate is $\beta(t)$, and the relative transmissibility of asymptomatic infected and exposed individuals compared to symptomatic infected individuals are $\theta_A$ and $\theta_E$, respectively. The average incubation period is $1/\delta$ days. Once the incubation period ends, a proportion $p$ of the exposed individuals become asymptomatic infected and the rest proportion $1-p$ become symptomatic infected. The disease induced death rate is $\mu(t)$. It takes an average of $1/r_I$ days and $1/r_A$ days for symptomatic and asymptomatic infected individuals to recover, respectively. The model is given by the following system of differential equations:

\begin{equation}
		\label{post_model}
		\aligned
		\frac{\mathrm{d}S(t)}{\mathrm{d}t} =& 	-\frac{\beta(t)S(t)(I(t)+\theta_EE(t)+\theta_AA(t))}{N}-\eta(t)S(t), \\
		\frac{\mathrm{d}E(t)}{\mathrm{d}t} =&	\frac{\beta(t)(S(t)+\epsilon_1V_1(t)+\epsilon_2V_2(t))(I(t)+\theta_EE(t)+\theta_AA(t))}{N}-\delta E(t), \\
		\frac{\mathrm{d}I(t)}{\mathrm{d}t} =& 
		(1-p)\delta E(t)-(\mu(t)+r_I)I(t), \\
		\frac{\mathrm{d}A(t)}{\mathrm{d}t} =& 
		p\delta E(t)-r_AA(t), \\
		\frac{\mathrm{d}R(t)}{\mathrm{d}t} =& 
		r_II(t)+r_AA(t), \\
		\frac{\mathrm{d}V_1(t)}{\mathrm{d}t} =& 
		\eta(t)S(t)-\gamma(t)V_1(t)-\frac{\epsilon_1\beta(t)V_1(t)(I(t)+\theta_EE(t)+\theta_AA(t))}{N}, \\
		\frac{\mathrm{d}V_2(t)}{\mathrm{d}t} =& 
		\gamma(t)V_1(t)-\frac{\epsilon_2\beta(t)V_2(t)(I(t)+\theta_EE(t)+\theta_AA(t))}{N}.
		\endaligned
\end{equation}

\subsection{Parameter estimation}

For the constant parameters $N$, $\delta$, $p$, $r_I$, $r_A$, we took the same values as those in our pre-vaccination model for the US \cite{Wang2021}; that is, $N=331,449,281$, $\delta=1/14$, $p=0.7$, $r_I=1/14$, $r_A=1/14$. The two new constant parameters $\epsilon_1$ and $\epsilon_2$ were estimated according to vaccine efficacy, which is generally reported as a relative risk reduction (RRR). Our method used the relative risk (RR), i.e., the ratio of attack rates with and without vaccination to get RRR, which equals 1–RR \cite{Piero2021}. The RRR of the Pfizer–BioNTech BNT162b2 mRNA vaccine beginning $7$ days after the first dose to before the second dose is $68.5\%$, and the RRR of Pfizer after $7$ days of the second dose is $94.8\%$ \cite{Danuta2021}. Since most people in the US take the Pfizer and Moderna vaccines which have similar efficacy \cite{Piero2021}, we use the RRR of first- and second-dose Pfizer vaccine to approximate the values of $\epsilon_1$ and $\epsilon_2$, which leads to $\epsilon_1=1-0.685=0.315$ and $\epsilon_2=1-0.948=0.052$.

The death rate on day $i$ (denoted by $\mu[i]$) is estimated using the following formula
$$\mu[i]=\frac{\# \text{new deaths on day}\ i}{\# \text{currently infected individuals on day}\ i}.$$

The first-dose vaccination rate on day $i$ (denoted by $\eta[i]$) is estimated by the following formula 
$$\eta[i]=\frac{\# \text{individuals who received their first dose vaccine on day}\ i}{\# \text{susceptible individuals on day}\ i},$$
where the number of susceptible individuals on day $i$ equals the total population $N$ minus the number of all infected individuals on and before day $i$ (regardless of whether recovered or not) and then minus the number of individuals who have been vaccinated before day $i$.

The second-dose vaccination rate on day $i$ (denoted by $\gamma[i]$) is given by
$$\gamma[i]=\frac{\# \text{individuals who received their second dose vaccine on day}\ i}{\# \text{individuals who are partially vaccinated before day}\ i}.$$

To estimate the time-varying transmission rate, we used the inverse method that we created in \cite{Wang2021} which is motivated by \cite{Kong2015,Pollicott2012}. We started by obtaining the time series of $E(t)$ from the term $(1-p)\delta E(t)$ which can be approximated by the notification data.
We use $S[k]$, $E[k]$, $I[k]$, $A[k]$, $R[k]$, $V_1[k]$, $V_2[k]$ to represent the values of variables in model \eqref{post_model}, $D[k]$ to represent the cumulative deaths, and $y[k]$ to be the notification data, on the $k$-th day of study. Then we have
$$E[k]=\frac{y[k]}{(1-p)\delta},\quad k=1,2,3,...,K,$$
where $K$ is the length of the vector of notification data.
We can obtain the time series data of $V_1[k]$ and $V_2[k]$, $k=1,2,3,...,K$, and the initial values $S[1]$, $I[1]$, $A[1]$, $R[1]$, $D[1]$ from reporting \cite{owidcoronavirus,worldometer}. 
It follows that

\begin{align*}
    I[i]&=I[i-1]+(1-p)\delta E[i-1]-(\mu[i-1]+r_I)I[i-1],\\
    A[i]&=A[i-1]+p\delta E[i-1]-r_A A[i-1],\\
    R[i]&=R[i-1]+r_I I[i-1]+r_A A[i-1],\\
    D[i]&=D[i-1]+\mu[i-1]I[i-1],\\
    S[i]&=N-E[i]-I[i]-A[i]-R[i]-D[i]-V_1[i]-V_2[i],
\end{align*}
for $i=2,3,...K$.

Next we add up the $S$-equation, $V_1$-equation and $V_2$-equation to obtain
$$\frac{\mathrm{d}(S(t)+V_1(t)+V_2(t))}{\mathrm{d}t}=-\frac{\beta(t)(S(t)+\epsilon_1V_1(t)+\epsilon_2V_2(t))(I(t)+\theta_EE(t)+\theta_AA(t))}{N}.$$
Substituting the time series of $S(t)$, $V_1(t)$, $V_2(t)$, $I(t)$, $E(t)$ and $A(t)$ into the above equation, we can solve for $\beta(t)$:
\begin{align*}
    \beta[i-1]=-\frac{N(S[i]+V_1[i]+V_2[i]-S[i-1]-V_1[i-1]-V_2[i-1])}{((S[i-1]+\epsilon_1V_1[i-1]+\epsilon_2V_2[i-1])(\theta_EE[i-1]+\theta_AA[i-1]+I[i-1]))}, \quad i=2,3,...,K,
\end{align*}
and we can approximate $\beta[K]$ by the value of $\beta[K-1]$.

\subsection{Prediction}

We performed three GBMs with different predictor variables to explore the relationship between the transmission rate and policy and/or mobility variables, and determine which factors mostly affect the transmission rate $\beta(t)$ according to their relative importance. Our main interest was in the GBM with all the 14 types of policy data (C1$\sim$C8, H1$\sim$H3, H6$\sim$H8) as predictor variables since the model with policy only has the power of prediction. The second GBM involves the mobility variables (M1$\sim$M6) only and it is used to explore the direct impact of human mobility on the prediction of the transmission rate. The last GBM incorporates all the mobility variables (M1$\sim$M6) together with four policy variables (H2, H3, H6, H7), aiming to investigate the joint impact of human mobility and policy on the prediction. Note that for the last GBM, we did not include the other policies since they are considered to have direct impact on human mobility and hence it may be unreasonable to put them together with the mobility as predictor variables when we want to change some of the policies whereas we do not know how the mobility varies accordingly.
By comparing the simulation results of these different GBMs, we were also able to see whether better predictions occur when mobility data are included.

The \texttt{gbm} package and \texttt{predict} function in \texttt{R} were used in the implementation of the gradient boosting machine learning. Since in the present study we focused on the effect of vaccination on the COVID-19 dynamics, we needed all the test durations to cover a part of post-vaccination period. To this end, we fixed the start data of training at April 4, 2020 and let the training duration vary from 231 days to 329 days by an increment of 7 days, and meanwhile fix each test duration at 35 days. Then the earliest test duration was from Nov 21, 2020 to Dec 25, 2020, which covers 6 days of the post-vaccination period, and the latest test duration consists of post-vaccination days from Feb 27, 2021 to April 2, 2021 as shown in Table \ref{postVac}.

The training dataset consisted of the transmission rate on each day obtained by the inverse method as the response variable and policy and/or mobility daily data as the predictor variables. For the first GBM, all the 14 types of policy data (C1$\sim$C8, H1$\sim$H3, H6$\sim$H8) on each day were provided as the predictor variables. For the second GBM, all the 6 types of mobility daily data (M1$\sim$M6) acted as the predictor variables. The training dataset of the third GBM contained all the mobility data (M1$\sim$M6) and the policy data H2, H3, H6, H7 as the predictor variables. 
For the testing dataset, we provided the trained GBMs with policy and/or mobility data to get the predicted transmission rate on each day of the test duration. 
Using the trained and tested transmission rate time series, we plotted the curve $(1-p)\delta E(t)$ of the SEIARVV model \eqref{post_model} and compared the simulated results with notification data of daily COVID-19 confirmed cases. 

The performance evaluation measures MAE (i.e., mean absolute error) and MAPE (i.e., mean absolute percentage error) were utilized to evaluate the differences between the predicted and actual numbers of confirmed cases, and the differences between the transmission rates predicted by GBMs and those derived from the inverse method.
The formulas of MAE and MAPE are given by
$$
\text{MAE}=\frac{1}{n}\sum_{i=1}^n|y_i-x_i|,\quad
\text{MAPE}=\frac{1}{n}\sum_{i=1}^n\left|\frac{y_i-x_i}{x_i}\right|,
$$
where $x_i$ is the $i$-th component of the vector of actual values, $y_i$ is the $i$-th component of the vector of prediction values, and
$n$ is the total number of data instances.
To obtain smaller MAE and MAPE, the GBMs were tuned with the number of trees, the distribution of response variable, the stochastic gradient descent, the depth of interaction, the learning rate, and the minimum number of observations allowed in the trees’ terminal nodes.

It is intriguing and important to know which predictors are more influential in training the GBM. We explored this by using the \texttt{summary} function in \texttt{R} which produces a table and a bar plot showing the values and ranking of the relative influence of each predictor variable. 

\section{Results}\label{Results}
\begin{figure}[ht]
\centering
\includegraphics[width=0.49\textwidth]{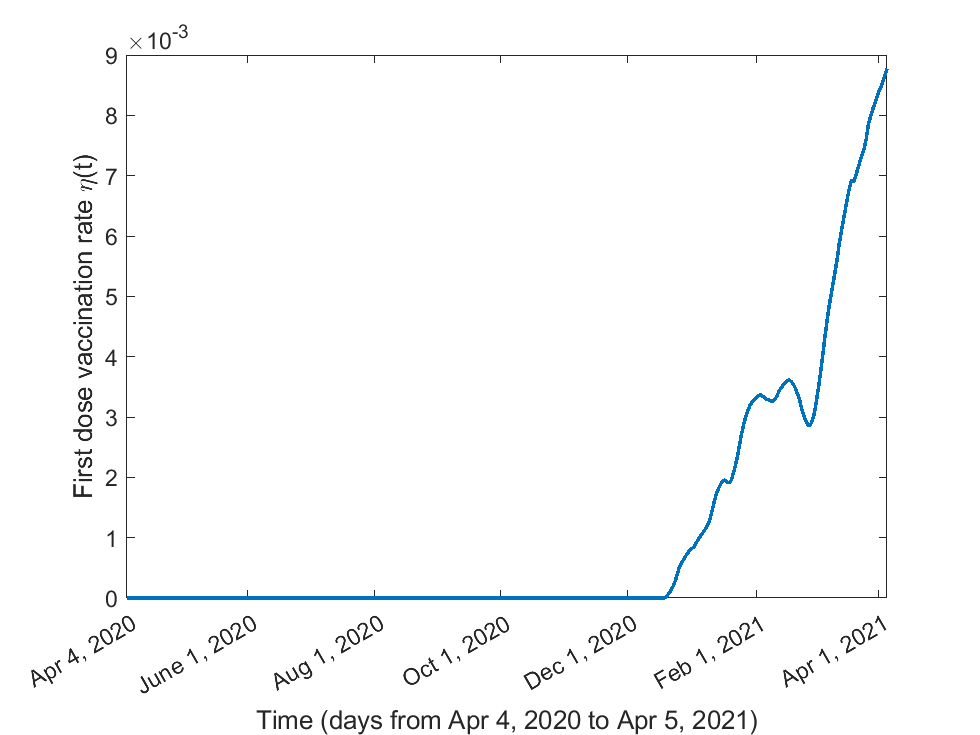}
\includegraphics[width=0.49\textwidth]{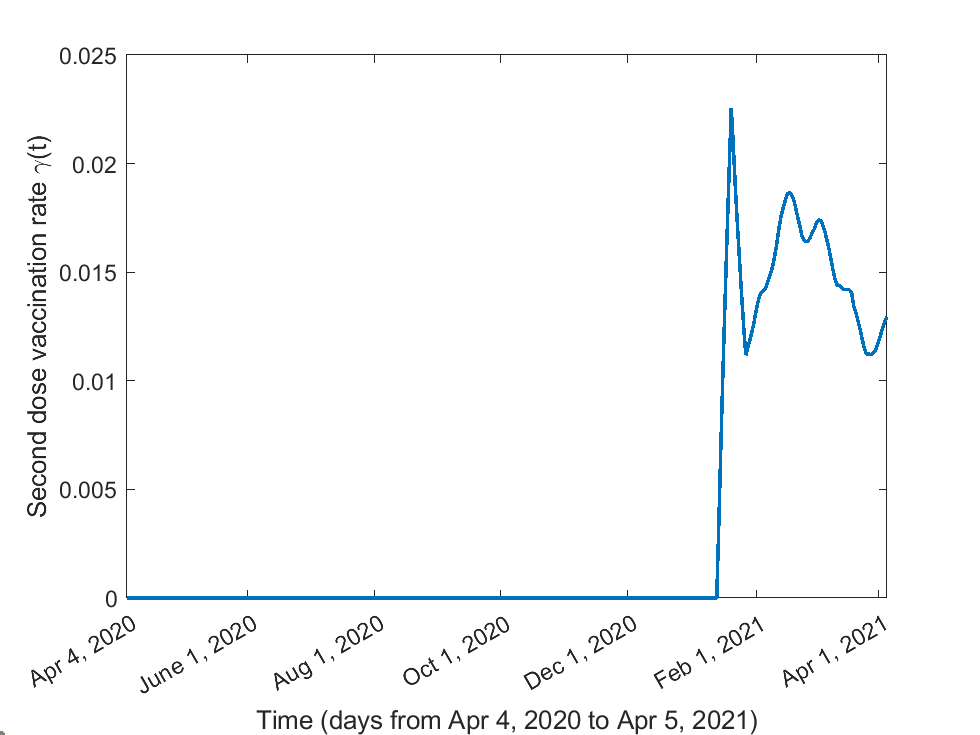}
\caption{\label{VacRate} First and second dose vaccination rates from Apr 4, 2020 to Apr 5, 2021.}
\end{figure}

\begin{figure}[ht]
\centering
\includegraphics[width=0.49\textwidth]{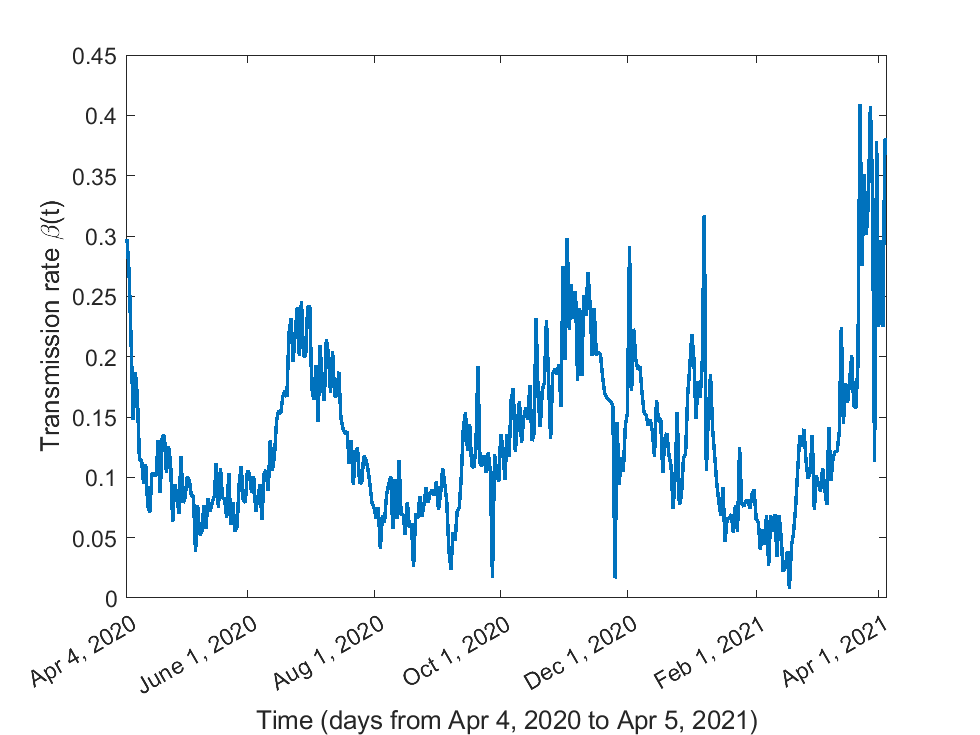}
\includegraphics[width=0.49\textwidth]{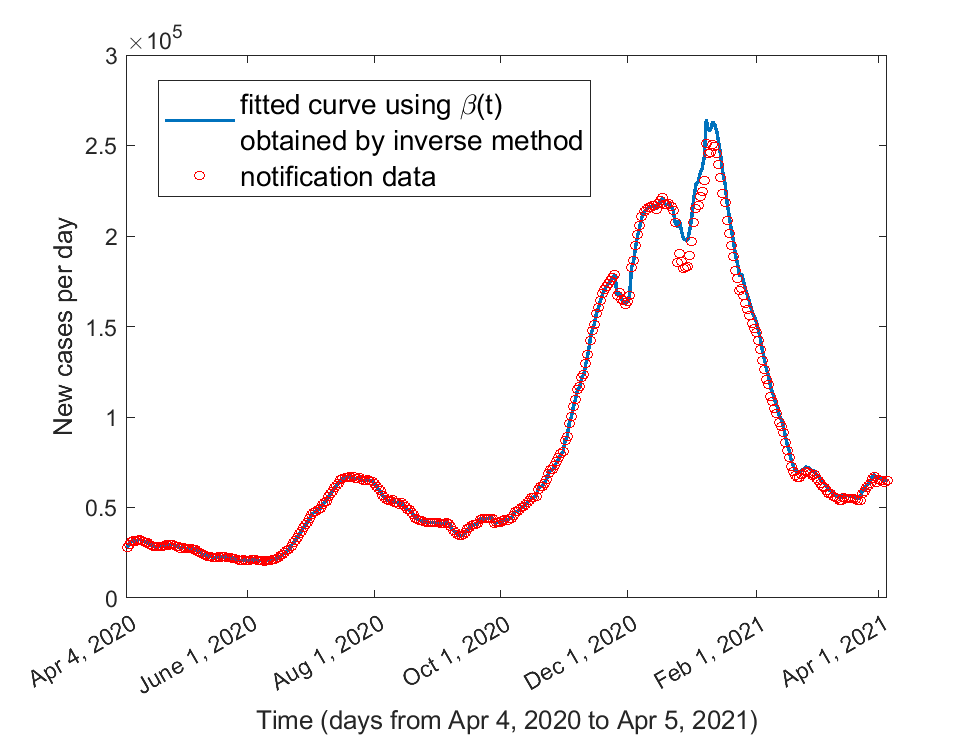}
\caption{\label{BetaPostVac} Transmission rate obtained by inverse method and the fitting with notification data from Apr 4, 2020 to Apr 5, 2021.}
\end{figure}

\begin{figure}[ht]
\centering
\includegraphics[width=0.49\textwidth]{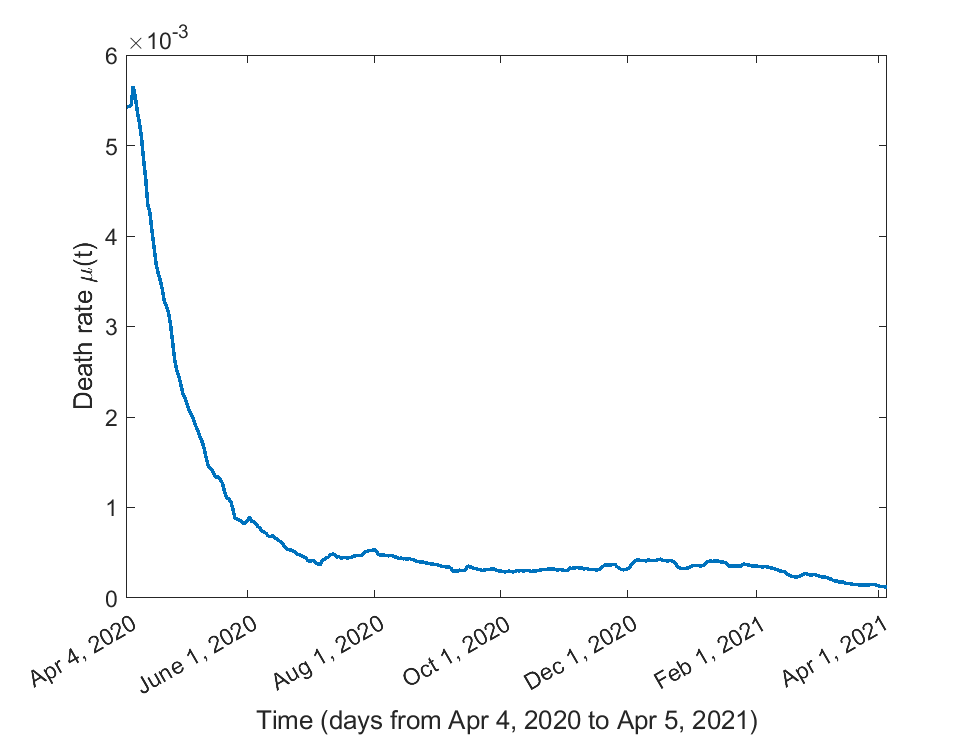}
\caption{\label{deathRate} Death rate of symptomatic infected individuals from Apr 4, 2020 to Apr 5, 2021.}
\end{figure}

The curves for the time-varying vaccination rates, the transmission rate obtained by the inverse method, and the death rate are shown in Figures \ref{VacRate}, \ref{BetaPostVac}, and \ref{deathRate}, respectively.
The GBMs based on the training and testing datasets perform better (i.e., lower MAE and MAPE) when the number of trees is $1000$, the stochastic gradient descent parameter is $0.9$, the depth of interaction is $30$, the learning rate is $0.01$, the minimum number of observations allowed in the trees’ terminal nodes is $10$, and the response variable has a Gaussian distribution.

\begin{figure}[ht]
\centering
\includegraphics[width=0.49\textwidth]{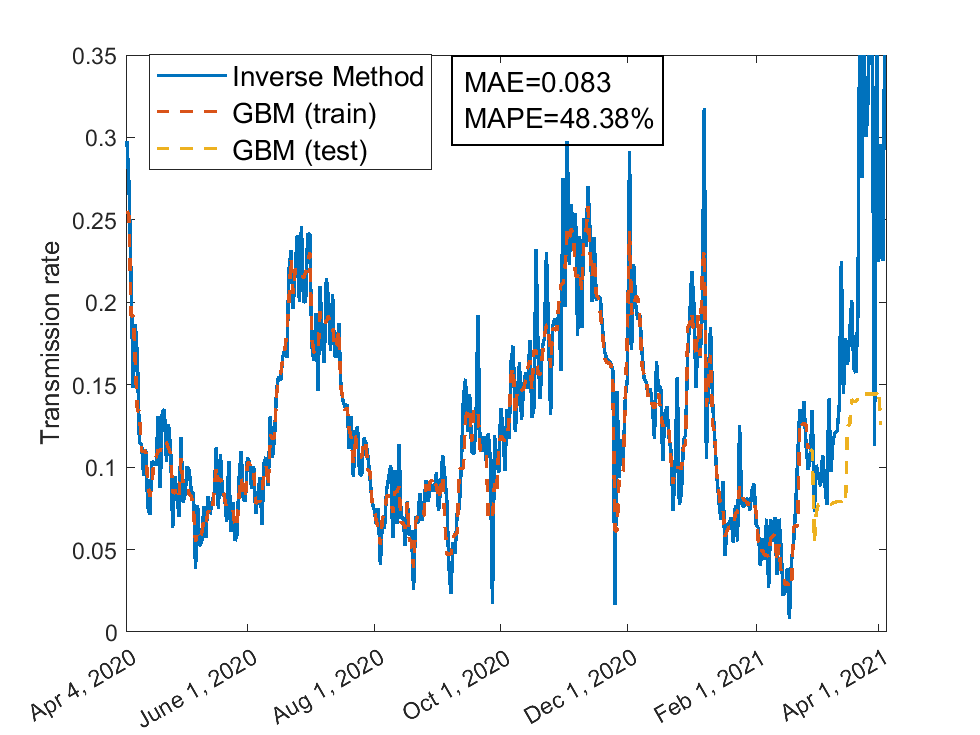}
\includegraphics[width=0.49\textwidth]{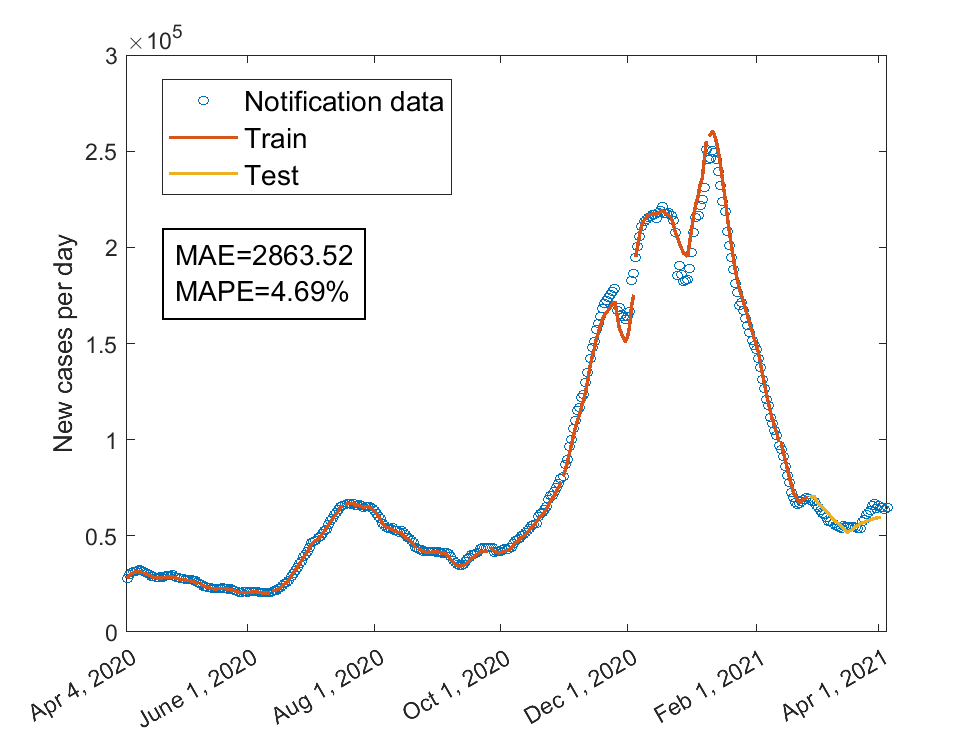}
\caption{\label{postBetaFit329P} Using policy data C1$\sim$C8, H1$\sim$H3, H6$\sim$H8, train 329 days from Apr 4, 2020 to Feb 26, 2021; test 35 days from Feb 27, 2021 to Apr 2, 2021.}
\end{figure}

\begin{figure}[ht]
\centering
\includegraphics[width=0.49\textwidth]{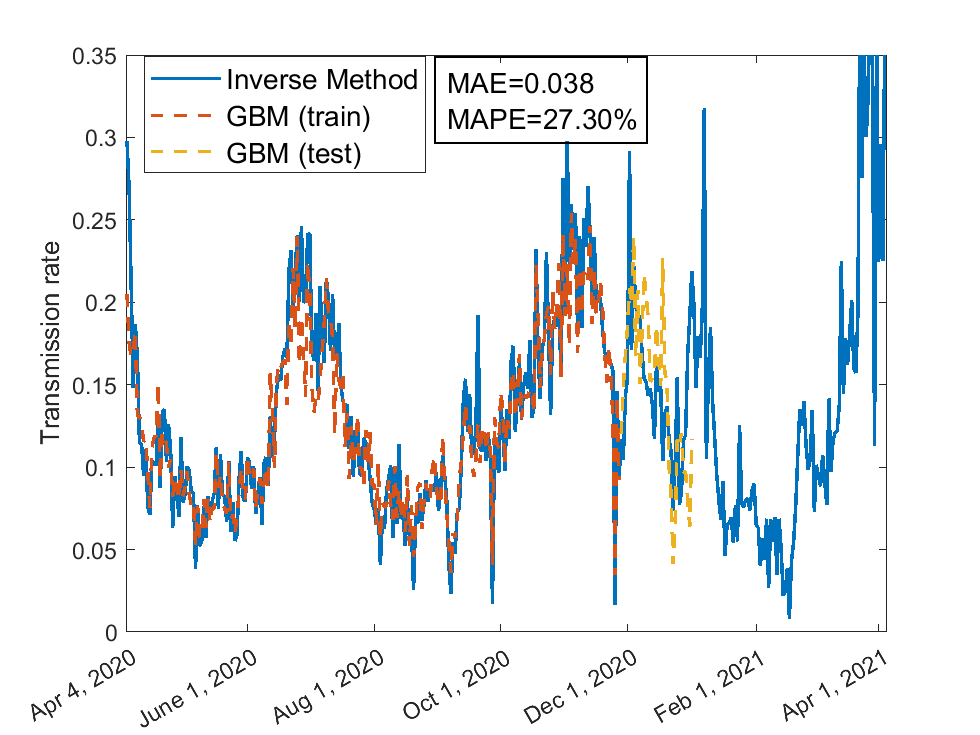}
\includegraphics[width=0.49\textwidth]{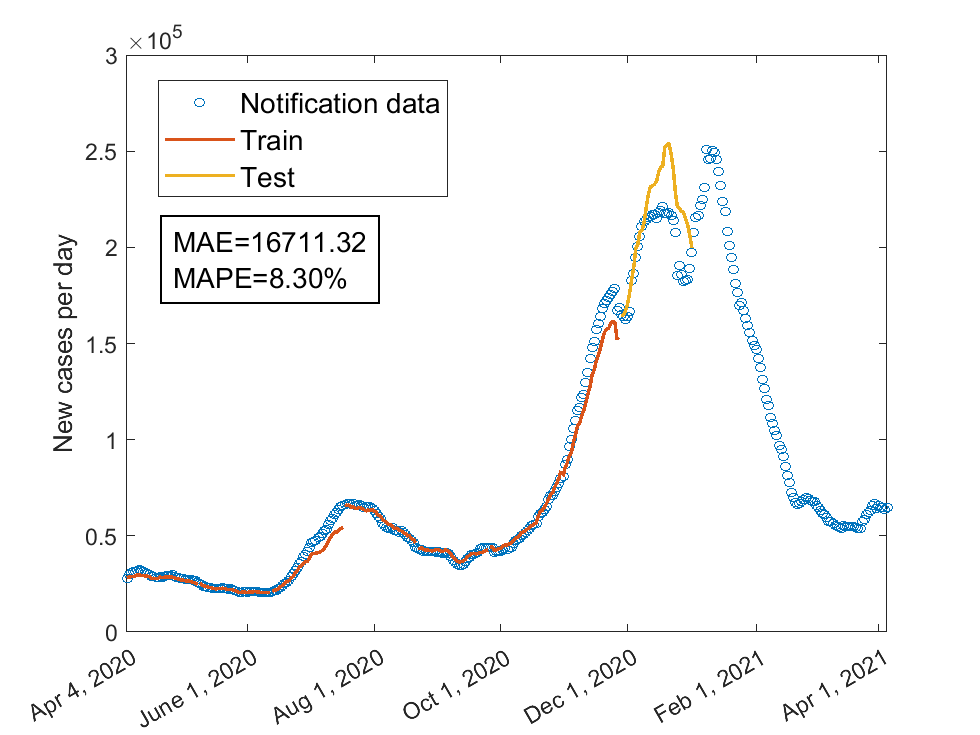}
\caption{\label{postBetaFit238M} Using mobility data M1$\sim$M6, train 238 days from Apr 4, 2020 to Nov 27, 2020; test 35 days from Nov 28, 2020 to Jan 1, 2021.}
\end{figure}

\begin{figure}[ht]
\centering
\includegraphics[width=0.49\textwidth]{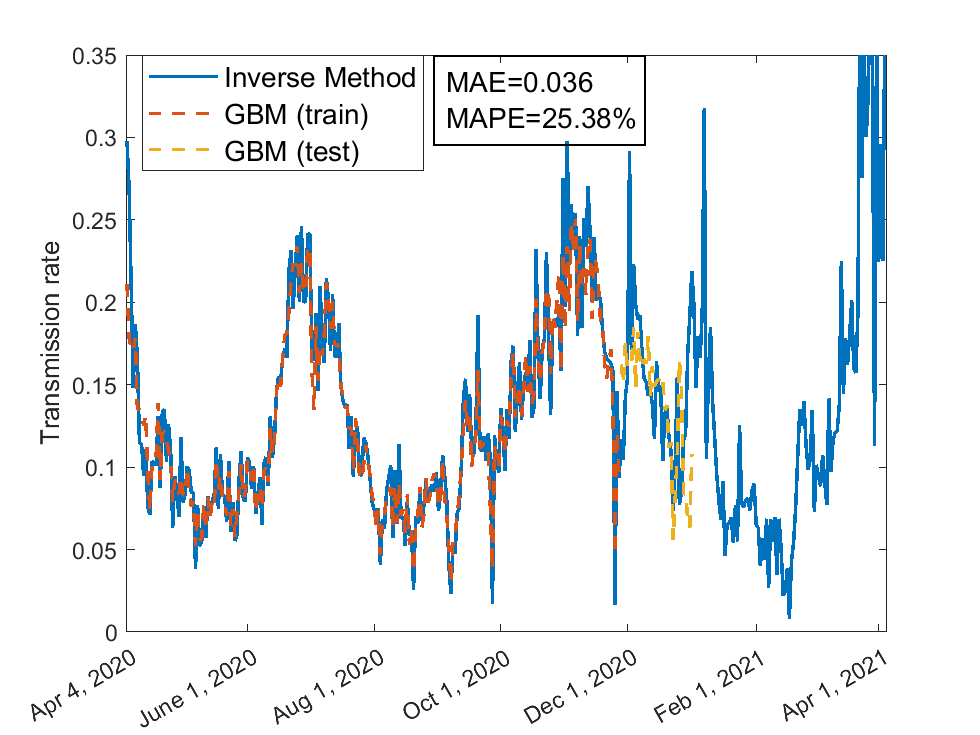}
\includegraphics[width=0.49\textwidth]{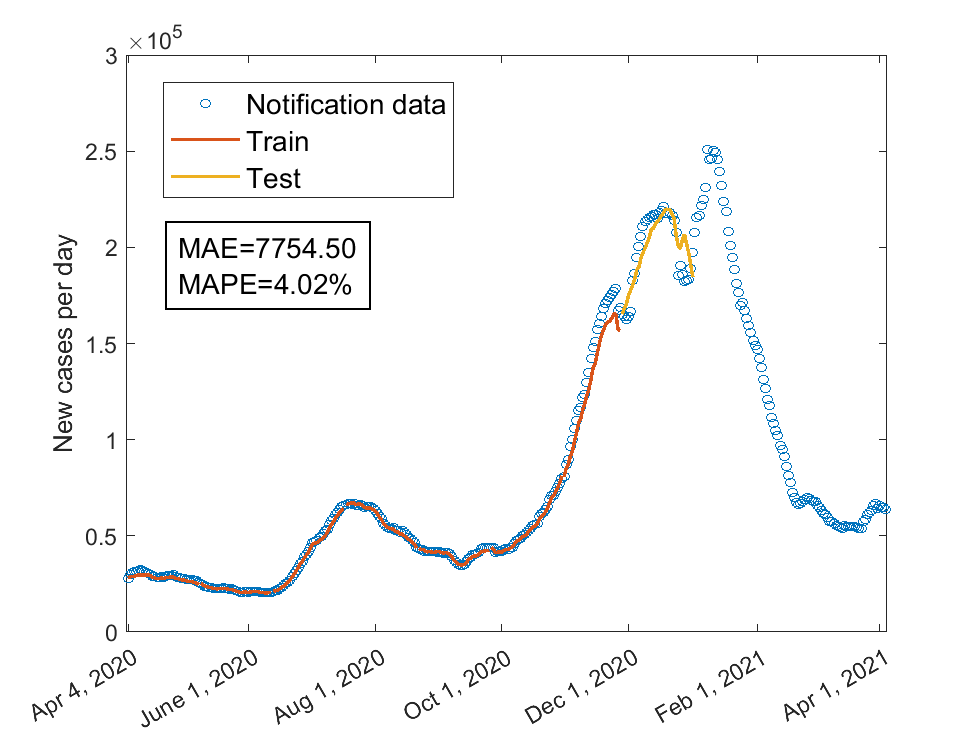}
\caption{\label{postBetaFit238MP7} Using mobility data M1$\sim$M6 and policy data H2, H3, H6, H7, train 238 days from Apr 4, 2020 to Nov 27, 2020; test 35 days from Nov 28, 2020 to Jan 1, 2021.}
\end{figure}

The averaged MAE and MAPE for the three GBMs across all different training durations are given in Table \ref{MAEtable2}. 
The MAE and MAPE corresponding to different training durations are presented in Table \ref{allMAEMAPE}.
The averaged MAE and MAPE of GBM with mobility only are much higher than the other two GBMs involving policies as predictor variables. 
The lowest averaged MAE and MAPE are obtained for the GBM which has both mobility and some of the policies as predictors.
However, as shown in Table \ref{allMAEMAPE}, the MAE and MAPE for the model with mobility only are not always the largest, and the MAE and MAPE for the model with both mobility and policy variables are not always the smallest for some specific training duration.
The best prediction result based on each GBM is shown in Figure \ref{postBetaFit329P} (MAPE=$4.69\%$), Figure \ref{postBetaFit238M} (MAPE=$8.30\%$), and Figure \ref{postBetaFit238MP7} (MAPE=$4.02\%$), respectively.
Some other selected training and testing results about the transmission rates as well as the fittings with notification data of daily confirmed cases corresponding to the training durations are presented in supplementary Figures \ref{postBetaFit252P}, \ref{postBetaFit259P} and \ref{postBetaFit308P} for the model with policy as the only predictors, in supplementary Figures 
\ref{postBetaFit245M} and \ref{postBetaFit273M} for the model with mobility as the only predictors, and in supplementary Figures \ref{postBetaFit245MP7}, 
\ref{postBetaFit252MP7}, \ref{postBetaFit294MP7},  \ref{postBetaFit308MP7} and \ref{postBetaFit315MP7} for the model with both mobility and policy as predictors (see Appendix).
We can see that the fittings with the transmission rate and the notification data of daily confirmed cases for the training part (orange curves in these figures) are almost perfect including the fitting with peaks and troughs.
Although the MAPEs for the predictions of the transmission rates are not small (greater than $25\%$), the MAPEs are quite small (most being smaller than $10\%$ ) for the predictions of notification data (see the yellow curves in these figures). 
In Figures \ref{postBetaFit238M}, \ref{postBetaFit238MP7}  and supplementary Figures  \ref{postBetaFit252P}, \ref{postBetaFit259P},  \ref{postBetaFit245M}, \ref{postBetaFit245MP7}, \ref{postBetaFit252MP7}, even if the training is based on pre-vaccination data, the predictions for post-vaccination confirmed cases have very small MAPEs: $8.30\%$, $4.02\%$, $6.87\%$, $9.07\%$,  $9.46\%$, $10.05\%$, $7.34\%$, respectively. 
In supplementary Figures \ref{postBetaFit252P} and \ref{postBetaFit259P}, although the MAPEs for the predictions of daily confirmed cases are small, the yellow curves in the right panels do not fit well with the local extremes of notification data. In contrast, when both policy and mobility variables are involved as predictors, the yellow curves in the right panels of Figure \ref{postBetaFit238MP7} fit well with the local extreme values of daily confirmed cases.
Besides, the yellow curves in the right panels of Figure \ref{postBetaFit238M} and supplementary Figures \ref{postBetaFit245M}, \ref{postBetaFit245MP7}, \ref{postBetaFit252MP7} replicate the trend of increasing to a local maximum and decreasing from the local maximum to a local minimum and then increasing to another peak. 

\begin{figure}[ht]
\centering
\includegraphics[width=0.6\textwidth]{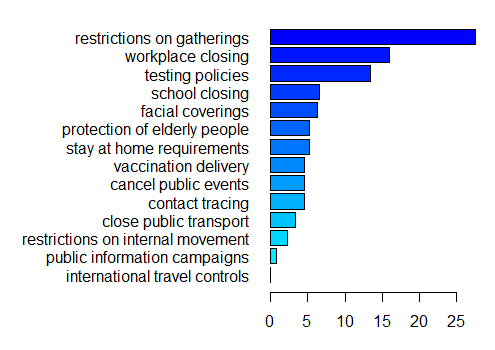}
\caption{\label{PostRI329P} Relative influence of policy variables C1$\sim$C8, H1$\sim$H3, H6$\sim$H8 when trained for 329 days from Apr 4, 2020 to Feb 26, 2021.}
\end{figure}

\begin{figure}[ht]
\centering
\includegraphics[width=0.5\textwidth]{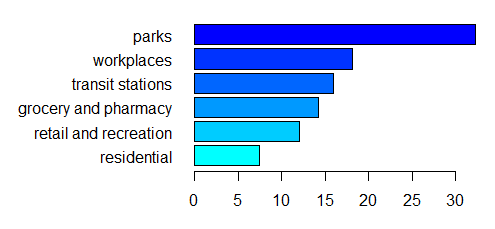}
\caption{\label{PostRI238M} Relative influence of mobility variables M1$\sim$M6 when trained for 238 days from Apr 4, 2020 to Nov 27, 2020.}
\end{figure}

\begin{figure}[ht]
\centering
\includegraphics[width=0.6\textwidth]{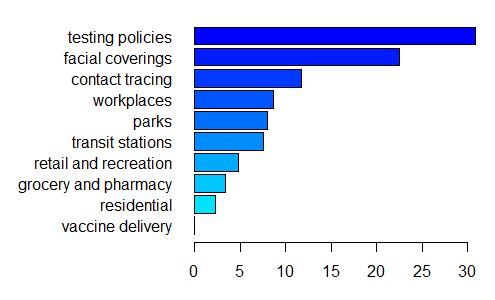}
\caption{\label{PostRI238MP7} Relative influence of mobility variables M1$\sim$M6 and policy variables H2, H3, H6, H7 when trained for 238 days from Apr 4, 2020 to Nov 27, 2020.}
\end{figure}

The relative influence of each policy variable in training the first GBM is shown in Table \ref{tablePost329P}, supplementary Tables \ref{tablePost252P}, \ref{tablePost259P}, \ref{tablePost308P}, 
and in Figure \ref{PostRI329P}, supplementary Figures \ref{PostRI252P}, \ref{PostRI259P}, \ref{PostRI308P} when the model is trained for 329 days, 252 days, 259 days, 308 days, respectively. 
From these tables and figures we can see that restrictions on gatherings is always the most influential policy regardless of how many days the model is trained for, which is consistent with the finding of our pre-vaccination paper \cite{Wang2021}.
Note that when trained for $252$ days and $259$ days, the relative influence of the vaccination delivery policy H7 is $0$ as these two training sets involve only pre-vaccination data. 
The vaccination delivery policy H7 becomes increasingly important as the training involves more post-vaccination data, with the relative influence increasing from $2.46\%$ when trained for $308$ days to $4.57\%$ when trained for $329$ days (see Table \ref{tablePost329P}, supplementary Table \ref{tablePost308P}, Figure \ref{PostRI329P} and supplementary Figure \ref{PostRI308P}).
The relative influence of the mobility variables in training the second GBM for 238 days, 245 days, and 273 days are given in Table \ref{tablePost238M}, supplementary Tables \ref{tablePost245M}, \ref{tablePost273M} and in Figure \ref{PostRI238M}, supplementary Figures \ref{PostRI245M}, \ref{PostRI273M}, respectively. 
For all these training sets, human mobility in parks is always the most important variable, followed by workplaces and transit stations, whereas residential mobility has the least influence on training the model. 
The relative influence of the mobility variables M1$\sim$M6 and the policies H2, H3, H6, H7 in training the third GBM, corresponding to training durations of 238 days, 245 days, 252 days, 294 days, 308 days, 315 days, are presented in Table \ref{tablePost238MP7}, supplementary Tables \ref{tablePost245MP7}, \ref{tablePost252MP7}, \ref{tablePost294MP7}, \ref{tablePost308MP7}, 
\ref{tablePost315MP7} and in Figure \ref{PostRI238MP7}, supplementary Figures
\ref{PostRI245MP7}, \ref{PostRI252MP7}, 
\ref{PostRI294MP7}, 
\ref{PostRI308MP7}, \ref{PostRI315MP7}, respectively.
When mobility data and part of the policy data are put together for training, the leading influential variable is always testing policy H2 with a weight of around $32\%$ in relative influence when the GBM is trained for 245 days. The second most important predictor is the facial covering policy H6 which weighs about $20\%$. The contact tracing policy H3 ranks in the third place, with a relative influence ranging from about $9.02\%$ to $15.38\%$. Similar to the GBM with policy predictors only, the ranking of vaccination delivery policy H7 increases when more post-vaccination days are included in the training dataset. The rankings of the mobility variables also change when the model is trained for different lengths of days.

\section{Discussion}\label{Discussion}
In this paper, we constructed a hybrid model by combining an ODE model with a variable transmission rate, motivated by the mechanisms of COVID-19 transmission and vaccination dynamics, coupled to a GBM, which provides a machine learning algorithm to forecast the transmission rate based on policy and/or mobility data. In our mechanistic model, we considered both partially vaccinated and fully vaccinated compartments in addition to the susceptible, exposed, symptomatic infected, asymptomatic infected and recovered compartments. In particular, we incorporated incidence terms in the vaccinated compartment equations to directly model the fact that vaccinated individuals can still get infected, which has rarely been studied by previous mathematical models with vaccination \cite{brett2020transmission,  macintyre2021modelling, li2020modeling}. The key step to link the ODE and GBM models is to obtain the time-varying transmission rate by the inverse method that we created in \cite{Wang2021}. This time-varying transmission rate can produce an almost perfect fit with the notification data of confirmed cases, which increases the chance of a good fitting using GBM. 
We trained the GBMs to fit the transmission rate obtained by the inverse method with policy and/or mobility data and predicted future transmission rate based on future policy and/or mobility data as well as the training experience. 
Then we used the trained and predicted transmission rate to plot solution curves of the mechanistic model to make predictions of the number of daily confirmed cases. 

The prediction performance was evaluated by mean absolute error (MAE) and mean absolute percentage error (MAPE). 
We found that the GBM trained on data on both mobility and some of the policies (testing, contact tracing, facial coverings, and vaccination delivery) is more efficient in establishing an association between the transmission rate and predictor variables than the GBM trained based on policy or mobility data only. 
The performance of the GBM with only mobility data performs the worst. 
Therefore, to model the impact of the preventive policies on the disease spread, mobility data appears to be insufficient.
Other factors, such as facial covering, must be included.

We also investigated the importance of the predictor variables and found that restrictions on gatherings is the most influential on training the GBM compared to other NPIs, which is consistent with its leading role in training the pre-vaccination model \cite{Wang2021}. This further emphasizes the importance of restrictions on gatherings even under vaccination. 
For the GBM with policy as the only predictor variables, testing policy is the second most influential predictor variable in most pre-vaccination cases (see \cite{Wang2021}). However, it is replaced by workplace closing or school closing in some post-vaccination cases. 
For the GBM with both policy and mobility as predictors, the rankings of the relative influence of the three policy variables testing, facial coverings and contact tracing are the same for both pre- and post-vaccinaion cases \cite{Wang2021}. For the GBM with mobility as the only predictors, parks and workplaces are always the two most influential factors regardless of the vaccination situation \cite{Wang2021}.
Since predictions are made based on training experience, investigation of relative influence of predictor variables in training the model can help us get closer to the mechanisms behind predictions.  
Although our machine-learning compartment is correlation based, which prevents causal statements on the relationship between gatherings restrictions and the transmission rate, our finding highlights the possibility of such causal relationship and motivates future work in this direction.
Indeed, some research works have already
tried to estimate the effects of different control measures or mobility on Covid-19 transmission dynamics. 
We would like to refer readers to \cite{Badr2020,Chinazzi2020,Xue2021,Lai2020,Joel2020} and the references therein.
Our relative influence results indicate that the vaccine delivery policy is not so 
important in training the model which seems different from the usual findings of most modeling works incorporating vaccination (see, e.g., \cite{macintyre2021modelling,Patel2021}). However, there is no disparity regarding this because the vaccine delivery policy in our model only describes the availability of vaccines such as to what extent or scale the vaccines are distributed to or donated to a region \cite{owidcoronavirus}. It may not represent how many people are actually getting vaccinated so it has no relation to either vaccination rate or cumulative vaccinated proportion.

For future work, it would be interesting to consider different infected compartments representing individuals infected with different strains of the virus. 
One can also study immunity waning cases where recovered and vaccinated individuals return to the susceptible compartment after a period of time. This aspect should be important in long-term forecasting. 
In addition to susceptible individuals, exposed and asymptomatic infected individuals who are unaware of their infections and recovered individuals can all get vaccinated, which will result in a new post-vaccination mechanistic model. 
It is informative to compare the COVID dynamics in different states of the US or other countries in the world.
The methods used in this work could be applied in the study of some other infectious disease transmission dynamics as well, especially when vaccination is implemented.

\section*{Acknowledgements}
This work was funded by Alberta Innovates and Pfizer via project number RES0052027.
HW gratefully acknowledges support from Natural Sciences and Engineering Research Council of Canada (NSERC) Discovery Grant RGPIN-2020-03911 and NSERC Accelerator Grant RGPAS-2020-00090. KN gratefully acknowledges support from National Institute for Mathematical Sciences (NIMS) grant funded by the Korean Government (NIMS-B21910000). 
MAL gratefully acknowledges a Canada Research Chair in Mathematical Biology and support from an NSERC Discovery Grant.

\bigskip

\begin{table}[ht]
\centering
\begin{tabular}{l|l|l}
Parameter & Interpretation & Value \\\hline
$\beta(t)$ & Transmission rate & see Figure \ref{BetaPostVac}  \\
$N$ & Total population of US & 331,449,281 \\
$\theta_E$ & Relative transmissibility of exposed individuals & 0.1\\ 
$\theta_A$ & Relative transmissibility of asymptomatic individuals  & 0.5 \\
$\eta(t)$  & First dose vaccination rate & see Figure \ref{VacRate} (a) \\
$\gamma(t)$ & Second dose vaccination rate & see Figure \ref{VacRate} (b) \\
$1/\delta$ & Incubation period &  14 days\\
$p$ & Proportion of asymptomatic infections & 0.7 \\
$\mu(t)$ & Death rate of symptomatic infected individuals & see Figure \ref{deathRate} \\
$r_I$ & Recovery rate of symptomatic infected individuals & 1/14 day$^{-1}$ \\
$r_A$ & Recovery rate of asymptomatic infected individuals & 1/14 day$^{-1}$\\
$\epsilon_1$ & Relative risk of infection for individuals having received & $0.315$ \\
&  only 1st-dose vaccine &  \\
$\epsilon_2$ & Relative risk of infection for fully vaccinated individuals & $0.052$  
\end{tabular}
\caption{\label{Parameters}Parameter interpretation and values.}
\end{table}

\begin{table}[ht]
\centering
\begin{tabular}{l|l|l}
Train length (days) & Train duration & Test duration \\\hline
 231
&Apr 4, 2020 to Nov 20, 2020
&Nov 21, 2020 to Dec 25, 2020\\
238
&Apr 4, 2020 to Nov 27, 2020
&Nov 28, 2020 to Jan 1, 2021\\
245
&Apr 4, 2020 to Dec 4, 2020
&Dec 5, 2020 to Jan 8, 2021\\
252
&Apr 4, 2020 to Dec 11, 2020
&Dec 12, 2020 to Jan 15, 2021\\
259
&Apr 4, 2020 to Dec 18, 2020
&Dec 19, 2020 to Jan 22, 2021\\
266
&Apr 4, 2020 to Dec 25, 2020
&Dec 26, 2020 to Jan 29, 2021\\
273
&Apr 4, 2020 to Jan 1, 2021
&Jan 2, 2021 to Feb 5, 2021\\
280
&Apr 4, 2020 to Jan 8, 2021
&Jan 9, 2021 to Feb 12, 2021\\
 287
&Apr 4, 2020 to Jan 15, 2021
&Jan 16, 2021 to Feb 19, 2021\\
294
&Apr 4, 2020 to Jan 22, 2021
&Jan 23, 2021 to Feb 26, 2021\\
301
&Apr 4, 2020 to Jan 29, 2021
&Jan 30, 2021 to Mar 5, 2021\\
308
&Apr 4, 2020 to Feb 5, 2021
&Feb 6, 2021 to Mar 12, 2021\\
315
&Apr 4, 2020 to Feb 12, 2021
&Feb 13, 2021 to Mar 19, 2021\\
322
&Apr 4, 2020 to Feb 19, 2021
&Feb 20, 2021 to Mar 26, 2021\\
329
&Apr 4, 2020 to Feb 26, 2021
&Feb 27, 2021 to April 2, 2021
\end{tabular}
\caption{\label{postVac}Training and testing durations.}
\end{table}

\begin{table}[htbp!]
\centering
\begin{tabular}{l|l|l}
Data used in GBM & Averaged MAE & Averaged MAPE \\\hline
Policy data C1$\sim$C8, H1$\sim$H3, H6$\sim$H8
&$42734.33$
&$34.41\%$\\
Mobility data M1$\sim$M6
&$42792.45$
&$45.23\%$\\
Mobility data M1$\sim$M6 and policy data H2,H3,H6,H7
&$26918.29$
&$21.09\%$
\end{tabular}
\caption{\label{MAEtable2}Averaged MAE and MAPE of the fittings with notification data.}
\end{table}

\begin{table}[ht]
\centering
\begin{tabular}{l|l|l|l|l|l|l}
Train length  & MAE  & MAPE ($\%$) & MAE  & MAPE ($\%$) & MAE  & MAPE ($\%$)\\
(days)& (GBM (1)) &(GBM (1)) & (GBM (2)) &(GBM (2)) & (GBM (3)) &(GBM (3))
\\\hline
231 &24101.06 & 12.30 & 31222.59 & 15.37 & 48502.27 & 24.14\\
238 & 72201.66 & 35.26 & 16711.32 & 8.30 & 7754.50 & 4.02\\
245 & 29735.10 & 14.74 & 19706.27 & 9.46 & 20422.00 & 10.05\\
252 & 13849.13 & 6.87 & 23758.06 & 10.78 & 15729.41 & 7.34\\
259 & 18213.25 & 9.07 & 32816.80 & 14.59 & 23390.42 & 10.42\\
266 & 31941.01 & 16.99 & 28850.16 & 13.18 & 26667.52 & 13.35\\
273 & 74313.63 & 47.23 & 19492.14 & 9.31 & 23146.67 & 12.74\\
280 & 130636.28 & 96.95 & 80407.18 & 57.72 & 102063.51 & 72.98\\
287 & 137085.86 & 127.97 & 70157.43 & 65.24 & 66216.15 & 59.70\\
294 & 45983.78 & 54.07 & 56215.69 & 66.92 & 10496.34 & 12.54\\
301 & 24090.06 & 30.78 & 76138.89 & 100.58 & 19361.91 & 24.30\\
308 & 6445.16 & 9.28 & 58948.11 & 89.23 & 7959.37 & 10.41\\
315 & 9929.69 & 16.41 & 49706.00 & 83.34 & 3907.50 & 5.66\\
322 & 19625.81 & 33.54 & 51335.51 & 88.74 & 16726.62 & 28.64\\
329 & 2863.52 & 4.69 & 26420.61 & 45.72 & 11430.18 & 20.06\\
\end{tabular}
\caption{\label{allMAEMAPE}MAE and MAPE of predictions of notification data based on model \eqref{post_model} and the three GBMs corresponding to different training durations. Predictors of GBM (1) are policy data C1$\sim$C8, H1$\sim$H3, H6$\sim$H8; predictors of GBM (2) are mobility data M1$\sim M6$; predictors of GBM (3) are  M1$\sim$M6 and policy data H2, H3, H6, H7.}
\end{table}

\begin{table}[ht]
\centering
\begin{tabular}{l|l}
Variable & Relative influence ($\%$)\\\hline
restrictions on gatherings &       $27.5967082$\\
workplace closing          &       $15.9506189$\\
testing policies           &       $13.4419815$\\
school closing             &        $6.6005727$\\
facial coverings           &        $6.2310945$\\
protection of elderly people &      $5.1828681$\\
stay at home requirements  &       $5.1504948$\\
vaccination delivery       &        $4.5671747$\\
cancel public events       &        $4.5633638$\\
contact tracing            &        $4.4840797$\\
close public transport     &        $3.2617845$\\
restrictions on internal movement & $2.1951767$\\
public information campaigns &      $0.7740819$\\
international travel controls &     $0.0000000$
\end{tabular}
\caption{\label{tablePost329P} Relative influence of policy variables C1$\sim$C8, H1$\sim$H3, H6$\sim$H8 when trained for 329 days from Apr 4, 2020 to Feb 26, 2021.}
\end{table}

\begin{table}[ht]
\centering
\begin{tabular}{l|l}
Variable & Relative influence ($\%$)\\\hline
parks & $32.283869$\\
workplaces & $18.126390$\\
transit stations & $15.907226$\\   
grocery and pharmacy & $14.162968$\\
retail and recreation & $12.037058$\\
residential & $7.482489$
\end{tabular}
\caption{\label{tablePost238M} Relative influence of mobility variables M1$\sim$M6 when trained for 238 days from Apr 4, 2020 to Nov 27, 2020.}
\end{table}

\begin{table}[ht]
\centering
\begin{tabular}{l|l}
Variable & Relative influence ($\%$)\\\hline
testing policies & $30.917619$\\         facial coverings & $22.572706$\\           contact tracing & $11.699315$\\              workplaces & $8.692511$\\
parks & $8.007173$\\    transit stations & $7.582054$\\ 
retail and recreation & $4.837882$\\ 
grocery and pharmacy &  $3.394418$\\               residential & $2.296323$\\ vaccine delivery & $0.000000$
\end{tabular}
\caption{\label{tablePost238MP7} Relative influence of mobility variables M1$\sim$M6 and policy variables H2, H3, H6, H7 when trained for 238 days from Apr 4, 2020 to Nov 27, 2020.}
\end{table}

\clearpage
\bibliographystyle{alpha}
\bibliography{sample}

\bigskip

\appendix

\section*{Appendix. Supplementary figures and tables}
\begin{figure}[ht]
\centering
\includegraphics[width=0.49\textwidth]{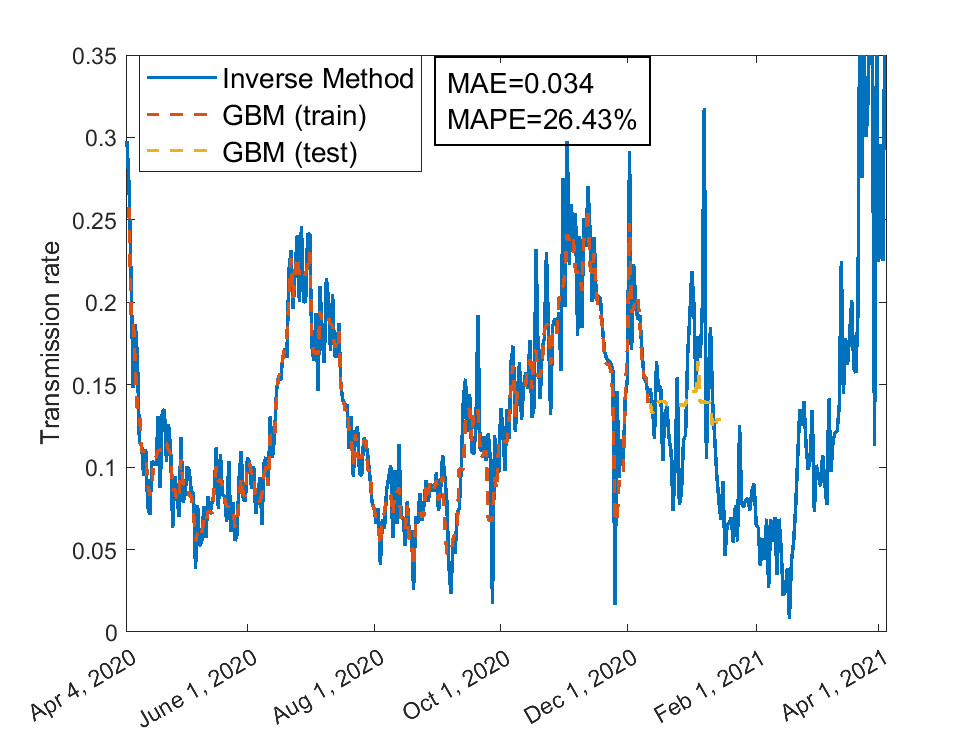}
\includegraphics[width=0.49\textwidth]{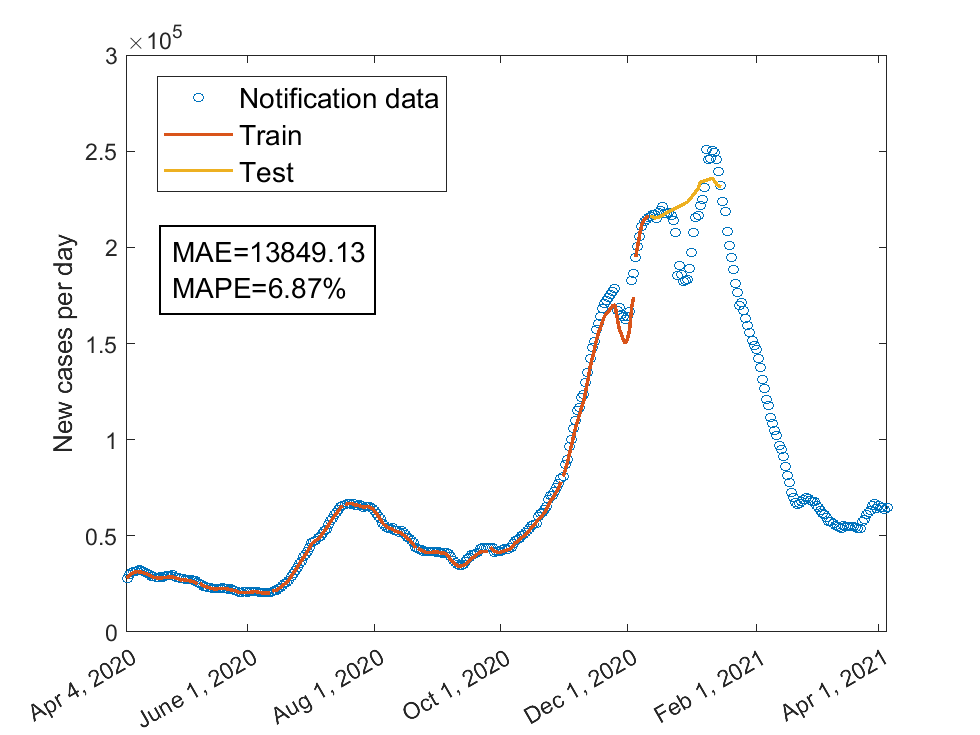}
\caption{\label{postBetaFit252P} Using policy data C1$\sim$C8, H1$\sim$H3, H6$\sim$H8, train 252 days from Apr 4, 2020 to Dec 11, 2020; test 35 days from Dec 12, 2020 to Jan 15, 2021.}
\end{figure}

\begin{figure}[ht]
\centering
\includegraphics[width=0.6\textwidth]{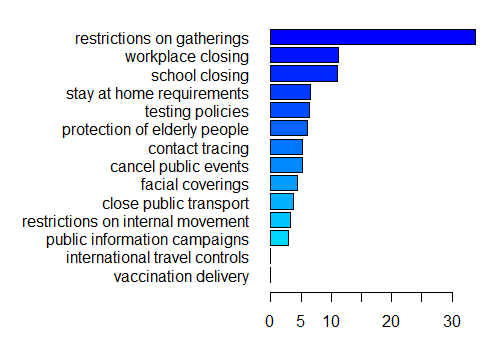}
\caption{\label{PostRI252P} Relative influence of policy variables C1$\sim$C8, H1$\sim$H3, H6$\sim$H8 when trained for 252 days from Apr 4, 2020 to Dec 11, 2020.}
\end{figure}

\begin{figure}[ht]
\centering
\includegraphics[width=0.49\textwidth]{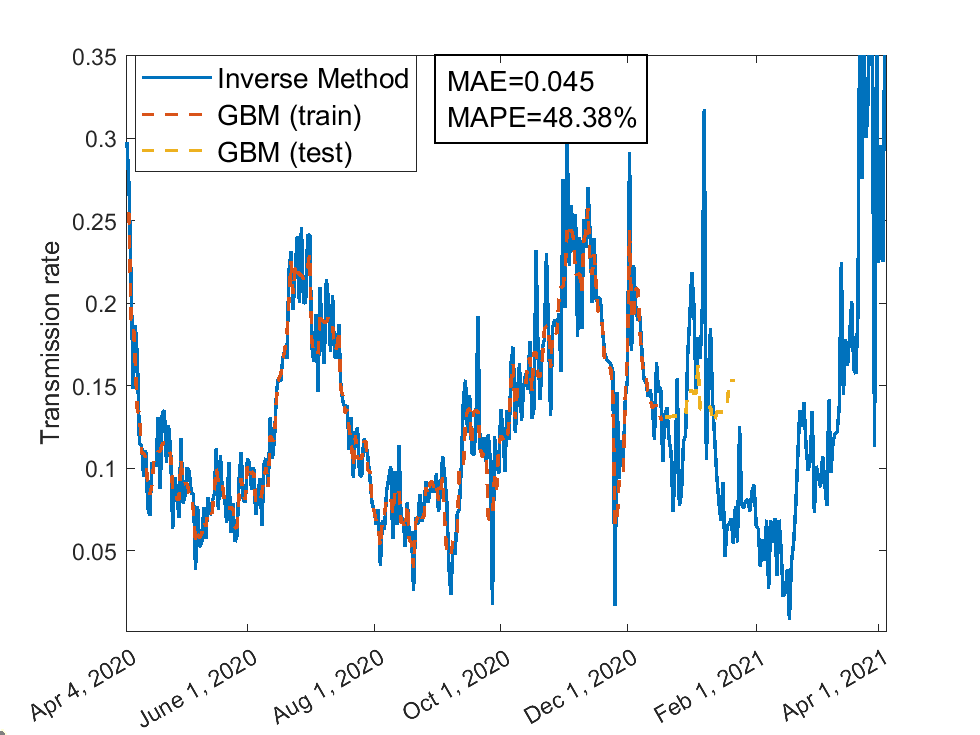}
\includegraphics[width=0.49\textwidth]{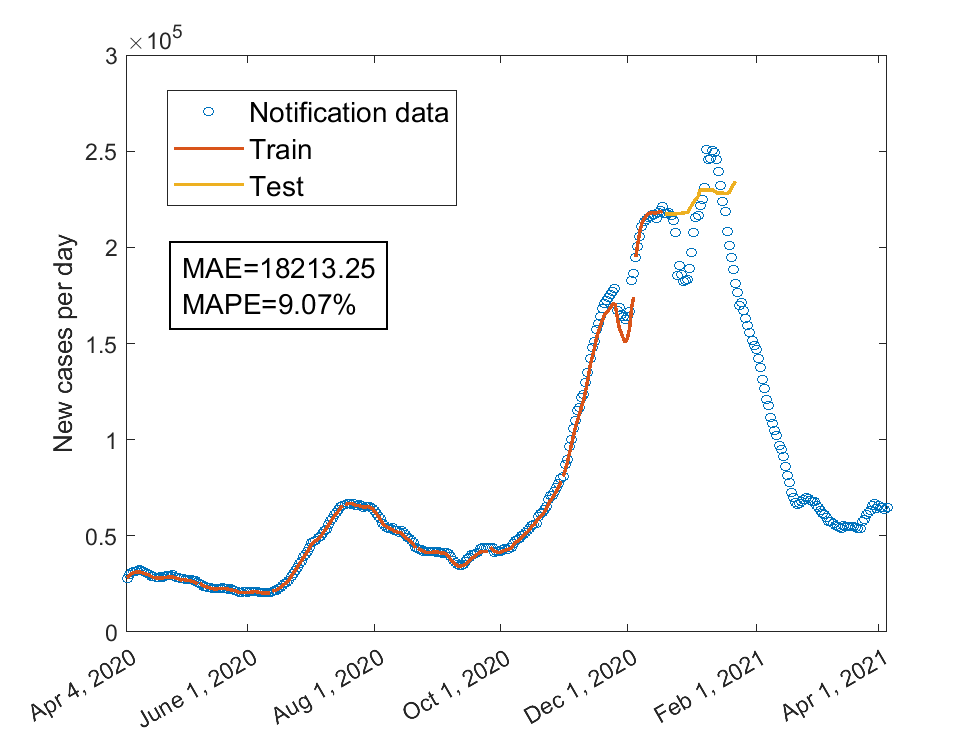}
\caption{\label{postBetaFit259P} Using policy data C1$\sim$C8, H1$\sim$H3, H6$\sim$H8, train 259 days from Apr 4, 2020 to Dec 18, 2020; test 35 days from Dec 19, 2020 to Jan 22, 2021.}
\end{figure}

\begin{figure}[ht]
\centering
\includegraphics[width=0.6\textwidth]{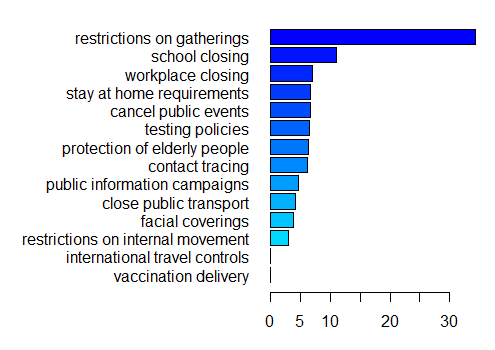}
\caption{\label{PostRI259P} Relative influence of policy variables C1$\sim$C8, H1$\sim$H3, H6$\sim$H8 when trained for 259 days from Apr 4, 2020 to Dec 18, 2020.}
\end{figure}

\begin{figure}[ht]
\centering
\includegraphics[width=0.49\textwidth]{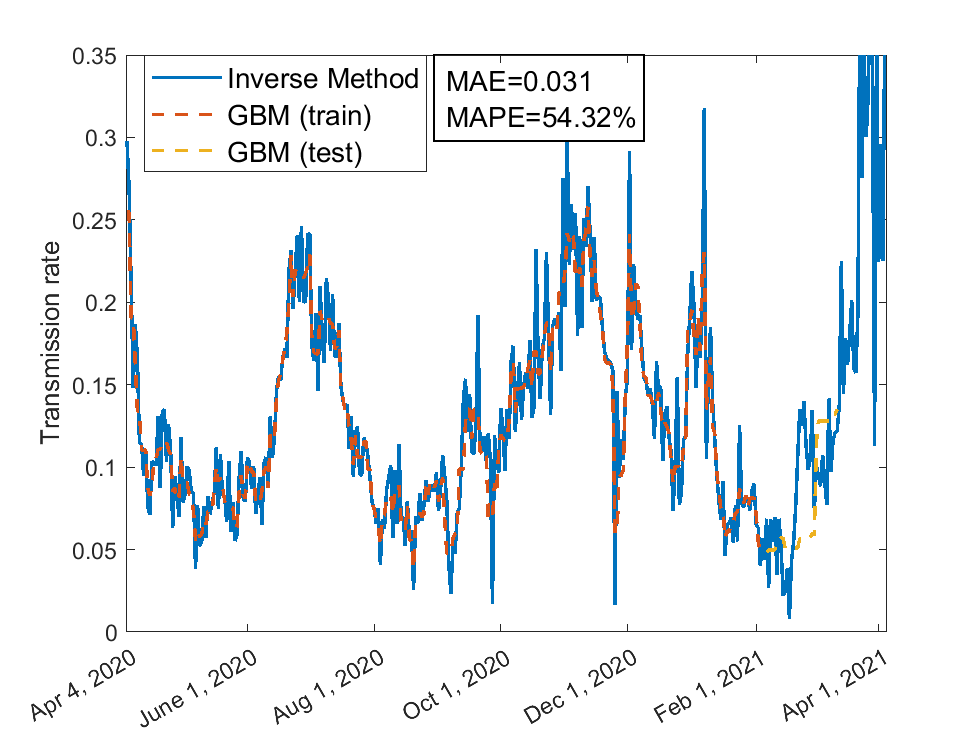}
\includegraphics[width=0.49\textwidth]{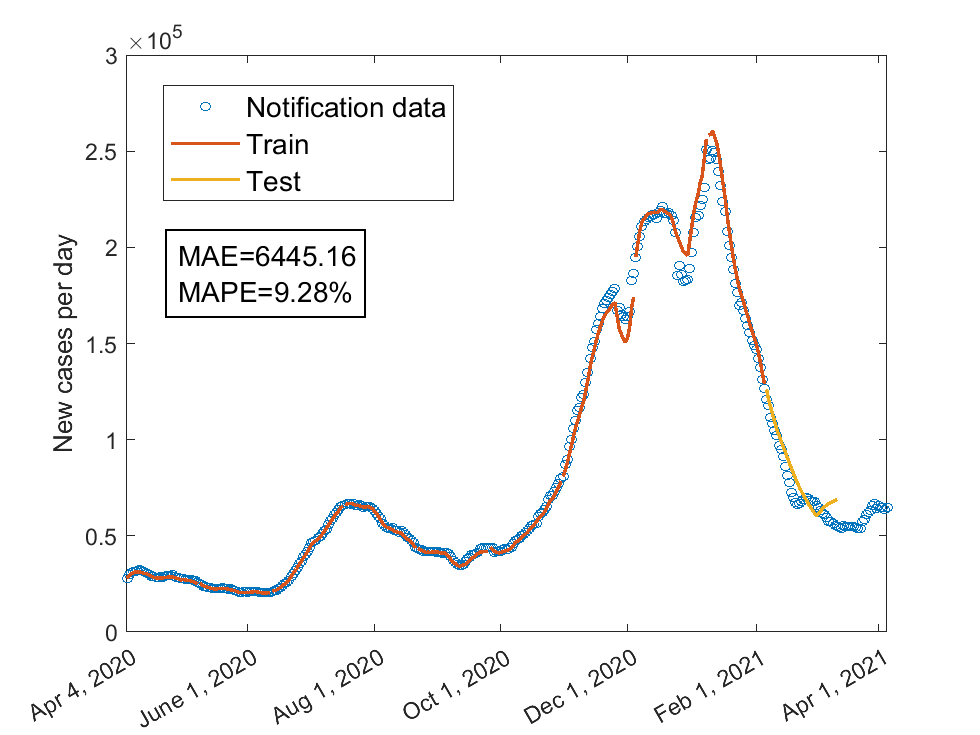}
\caption{\label{postBetaFit308P} Using policy data C1$\sim$C8, H1$\sim$H3, H6$\sim$H8, train 308 days from Apr 4, 2020 to Feb 5, 2021; test 35 days from Feb 6, 2021 to Mar 12, 2021.}
\end{figure}

\begin{figure}[ht]
\centering
\includegraphics[width=0.6\textwidth]{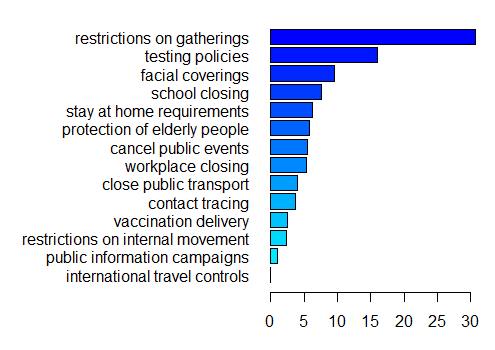}
\caption{\label{PostRI308P} Relative influence of policy variables C1$\sim$C8, H1$\sim$H3, H6$\sim$H8 when trained for 308 days from Apr 4, 2020 to Feb 5, 2021.}
\end{figure}

\begin{figure}[ht]
\centering
\includegraphics[width=0.49\textwidth]{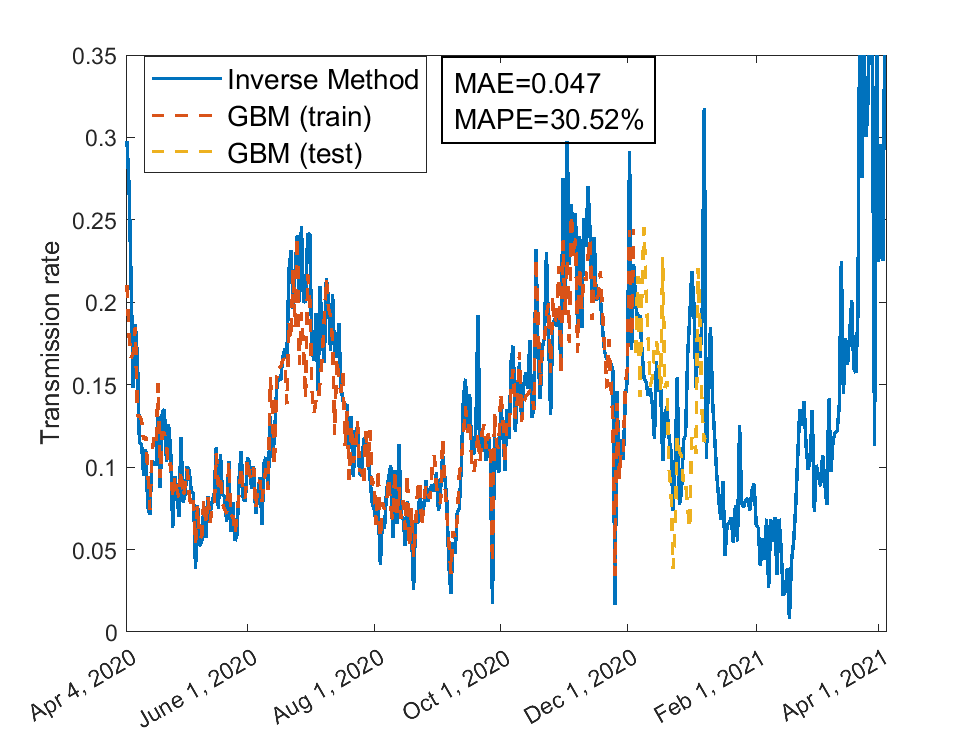}
\includegraphics[width=0.49\textwidth]{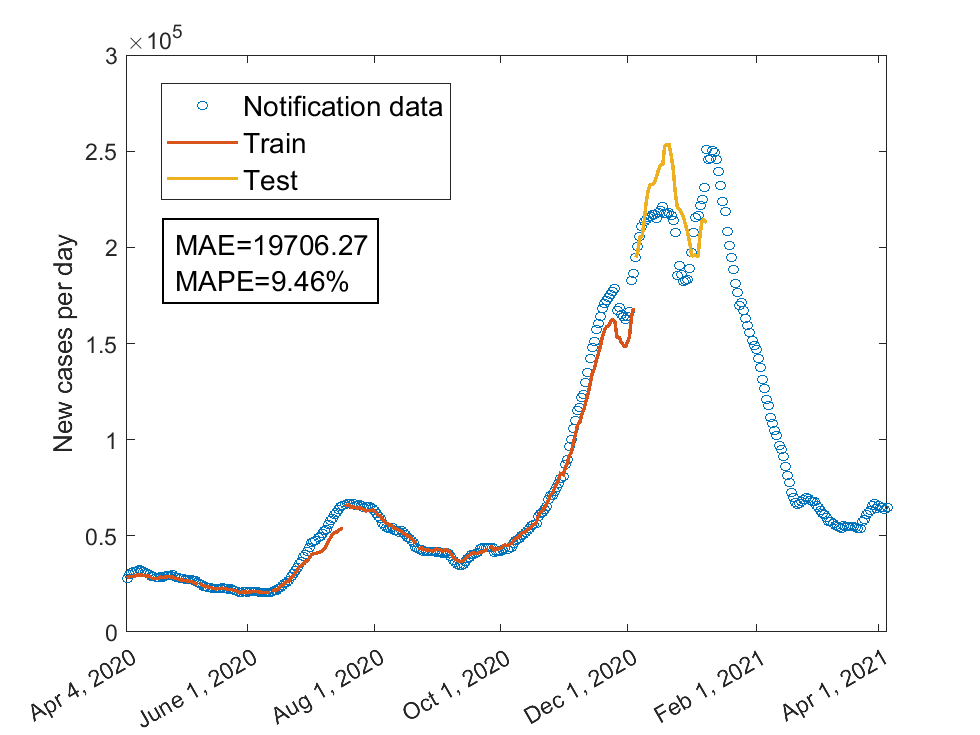}
\caption{\label{postBetaFit245M} Using mobility data M1$\sim$M6, train 245 days from Apr 4, 2020 to Dec 4, 2020; test 35 days from Dec 5, 2020 to Jan 8, 2021.}
\end{figure}

\begin{figure}[ht]
\centering
\includegraphics[width=0.5\textwidth]{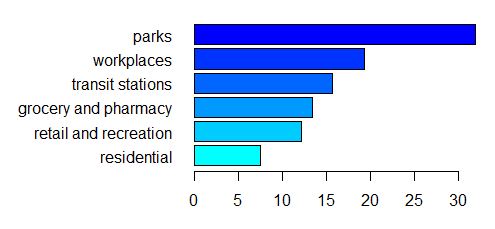}
\caption{\label{PostRI245M} Relative influence of mobility variables M1$\sim$M6 when trained for 245 days from Apr 4, 2020 to Dec 4, 2020.}
\end{figure}

\begin{figure}[ht]
\centering
\includegraphics[width=0.49\textwidth]{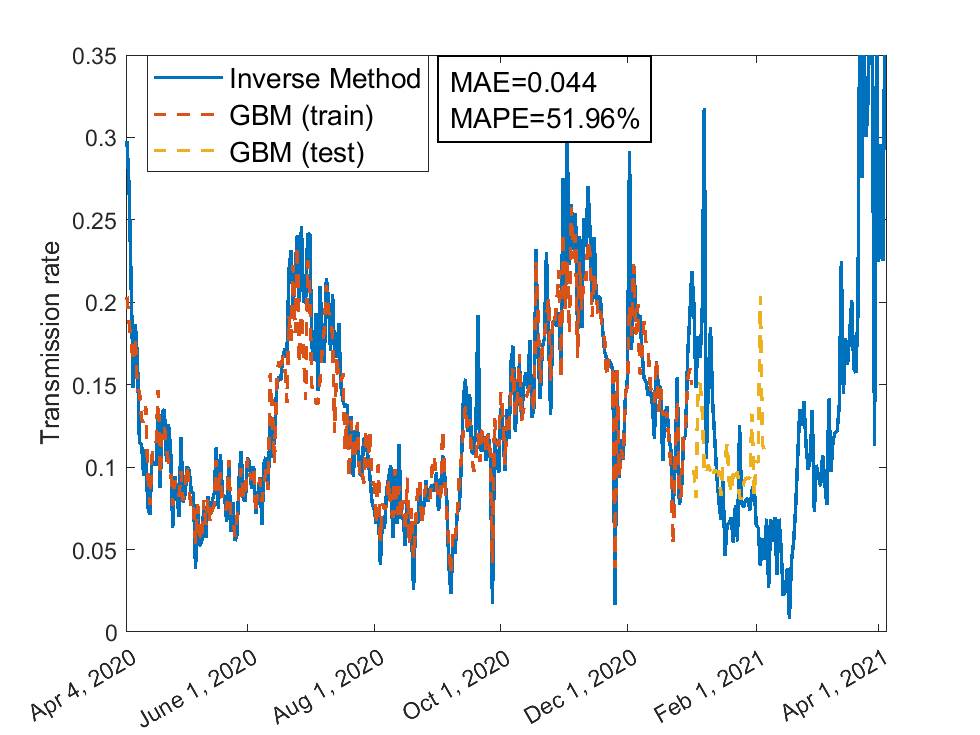}
\includegraphics[width=0.49\textwidth]{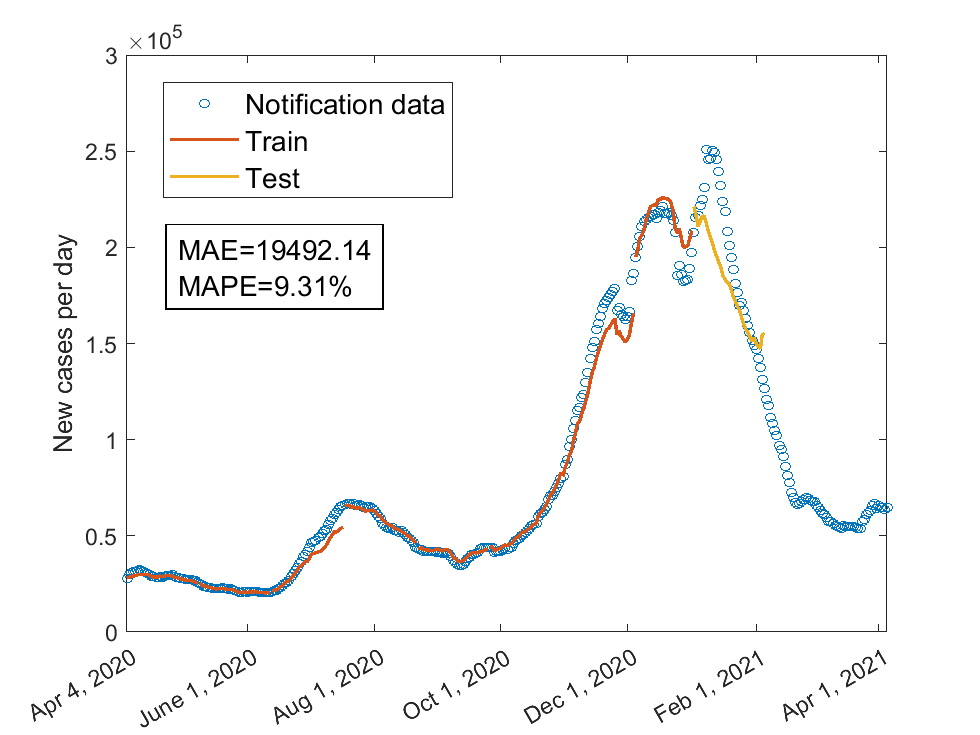}
\caption{\label{postBetaFit273M} Using mobility data M1$\sim$M6, train 273 days from Apr 4, 2020 to Jan 1, 2021; test 35 days from Jan 2, 2021 to Feb 5, 2021.}
\end{figure}

\begin{figure}[ht]
\centering
\includegraphics[width=0.5\textwidth]{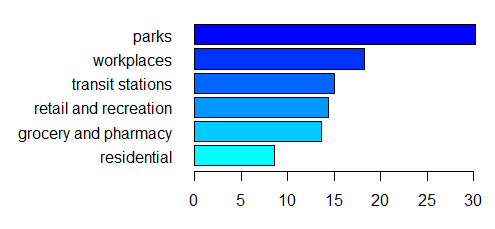}
\caption{\label{PostRI273M} Relative influence of mobility variables M1$\sim$M6 when trained for 273 days from Apr 4, 2020 to Jan 1, 2021.}
\end{figure}

\begin{figure}[ht]
\centering
\includegraphics[width=0.49\textwidth]{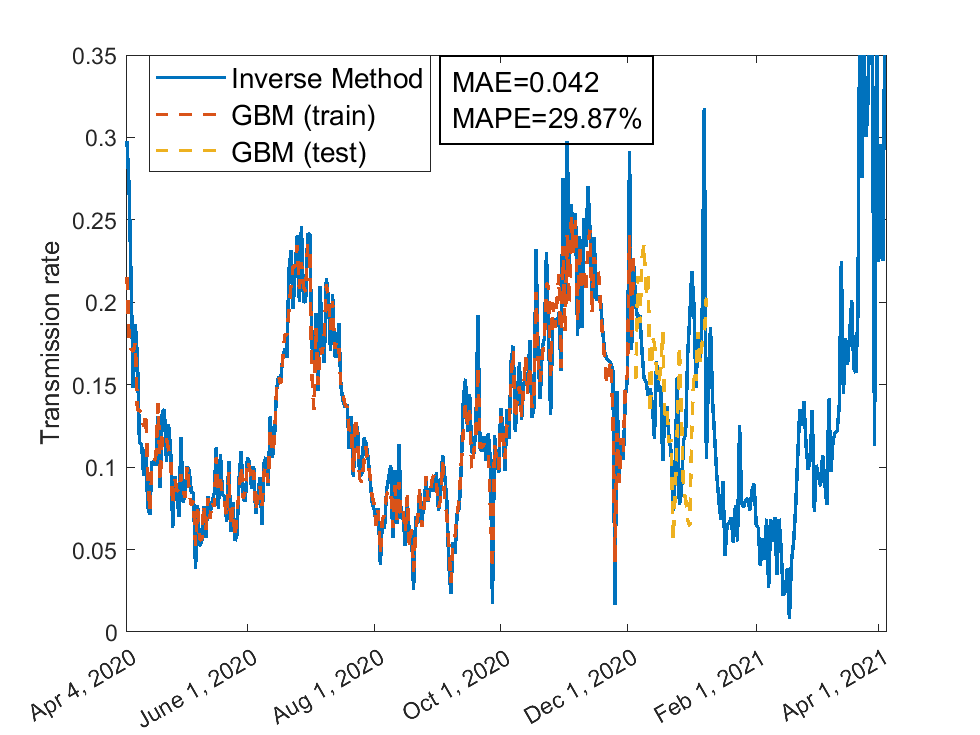}
\includegraphics[width=0.49\textwidth]{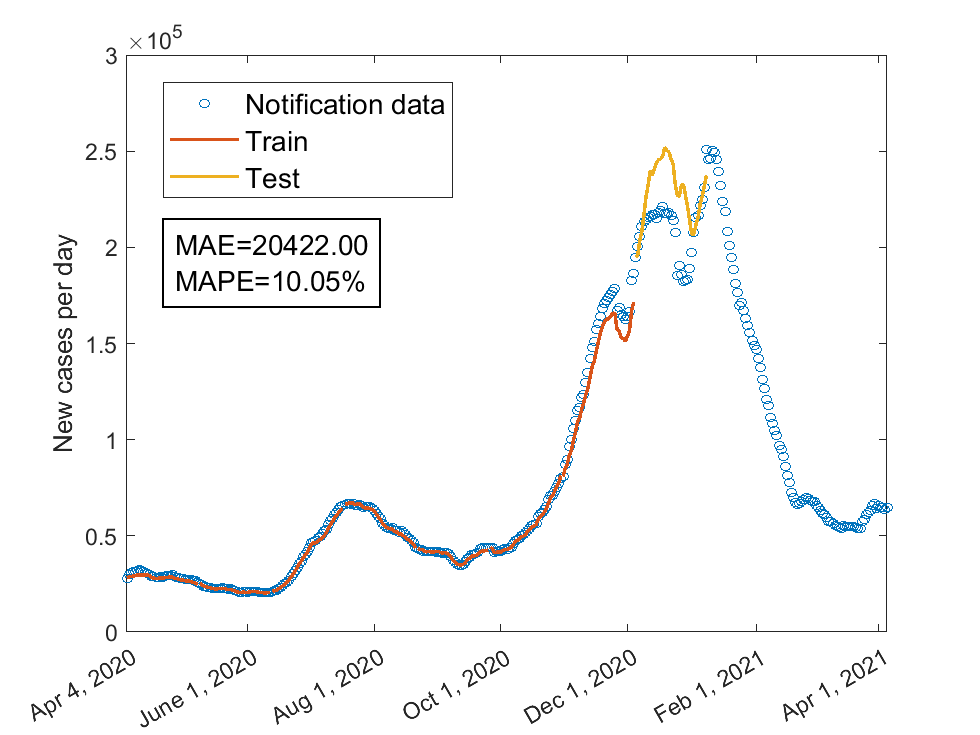}
\caption{\label{postBetaFit245MP7} Using mobility data M1$\sim$M6 and policy data H2, H3, H6, H7, train 245 days from Apr 4, 2020 to Dec 4, 2020; test 35 days from Dec 5, 2020 to Jan 8, 2021.}
\end{figure}

\begin{figure}[ht]
\centering
\includegraphics[width=0.6\textwidth]{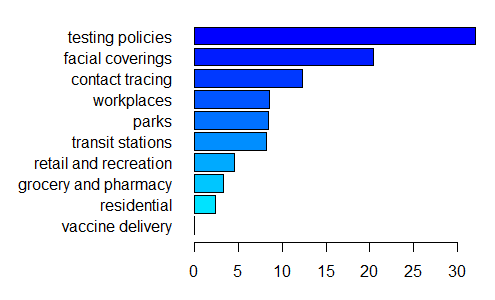}
\caption{\label{PostRI245MP7} Relative influence of mobility variables M1$\sim$M6 and policy variables H2, H3, H6, H7 when trained for 245 days from Apr 4, 2020 to Dec 4, 2020.}
\end{figure}

\begin{figure}[ht]
\centering
\includegraphics[width=0.49\textwidth]{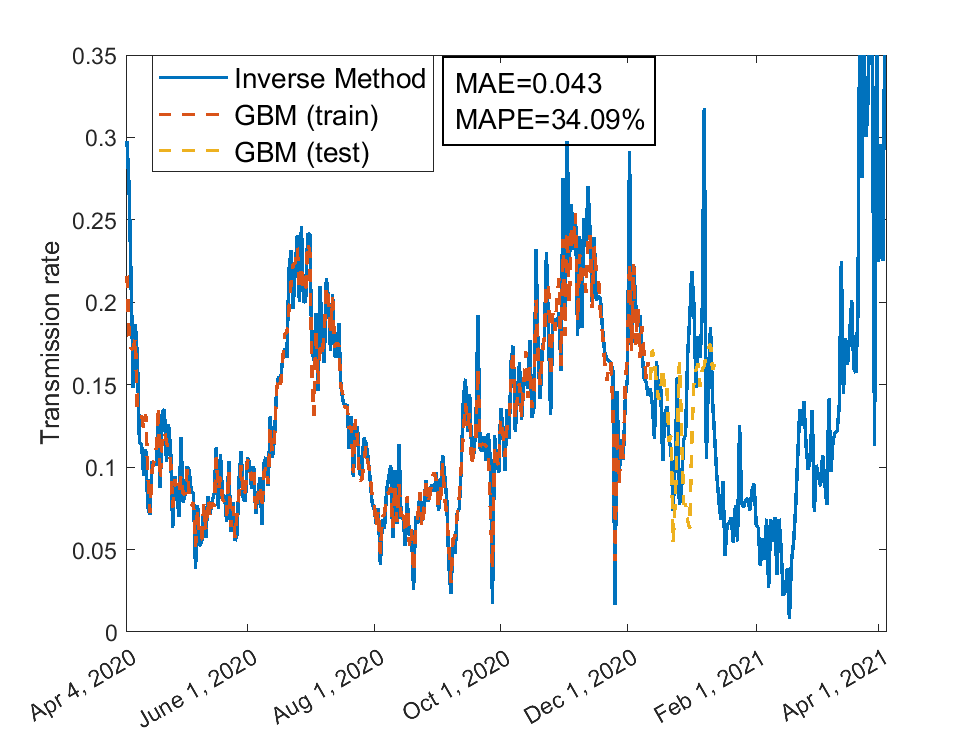}
\includegraphics[width=0.49\textwidth]{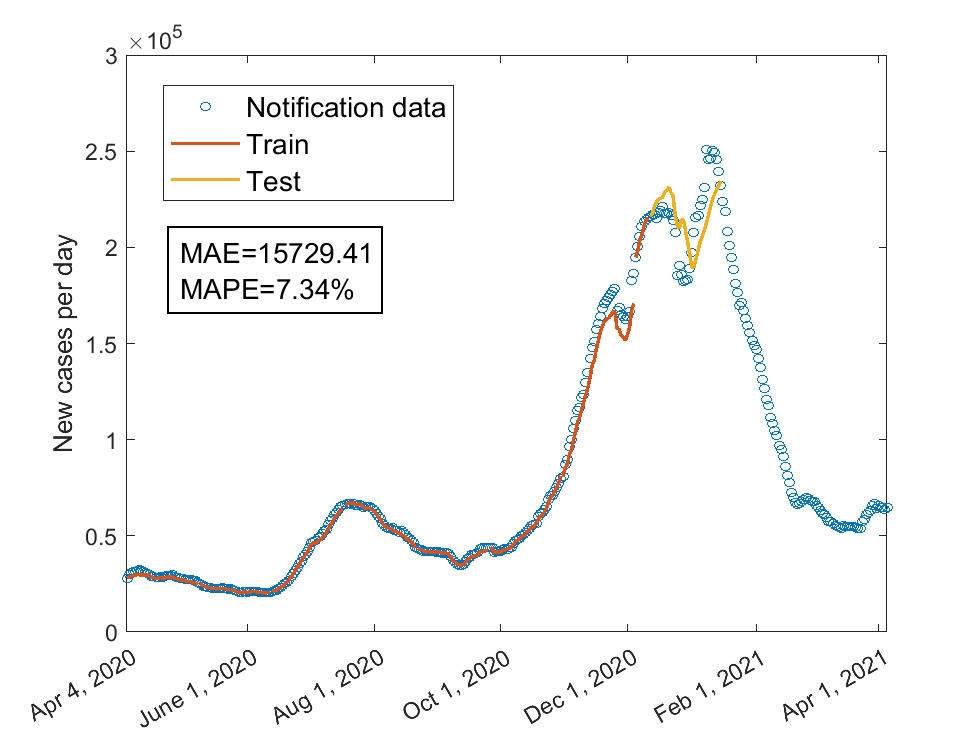}
\caption{\label{postBetaFit252MP7} Using mobility data M1$\sim$M6 and policy data H2, H3, H6, H7, train 252 days from Apr 4, 2020 to Dec 11, 2020; test 35 days from Dec 12, 2020 to Jan 15, 2021.}
\end{figure}

\begin{figure}[ht]
\centering
\includegraphics[width=0.6\textwidth]{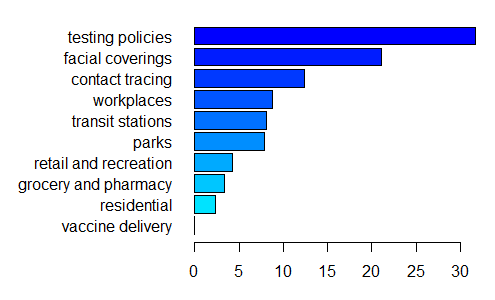}
\caption{\label{PostRI252MP7} Relative influence of mobility variables M1$\sim$M6 and policy variables H2, H3, H6, H7 when trained for 252 days from Apr 4, 2020 to Dec 11, 2020.}
\end{figure}

\begin{figure}[ht]
\centering
\includegraphics[width=0.49\textwidth]{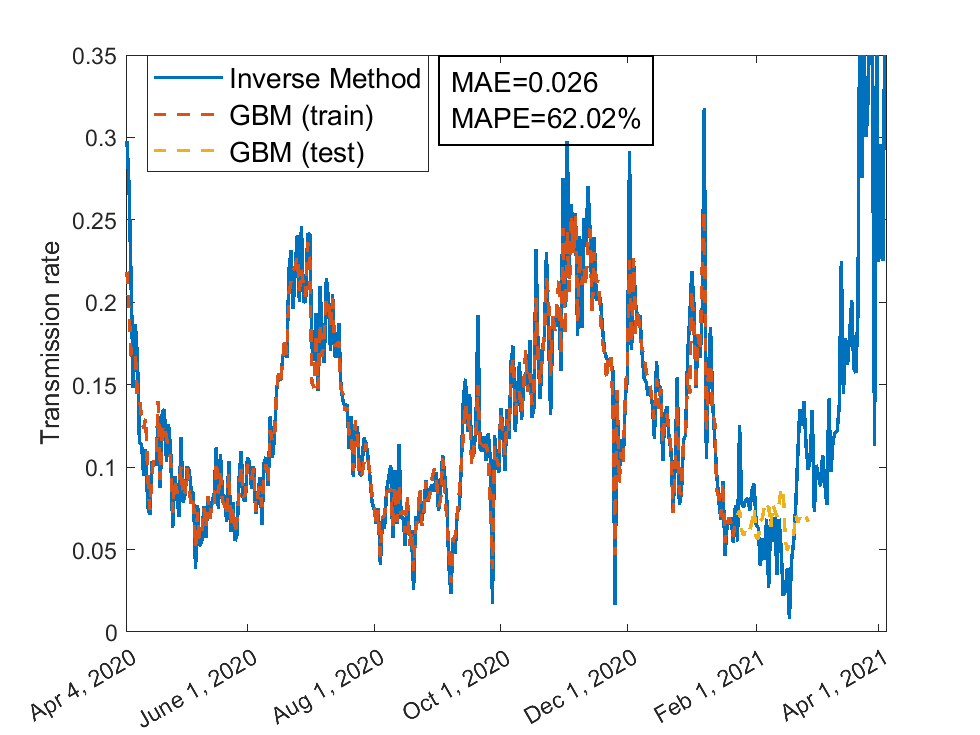}
\includegraphics[width=0.49\textwidth]{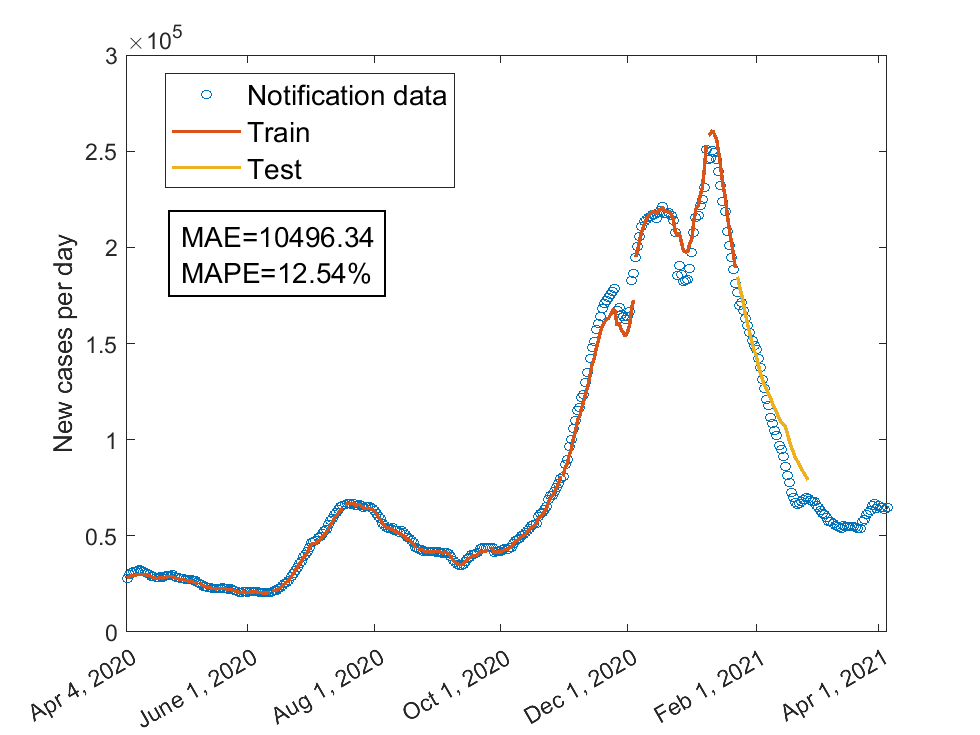}
\caption{\label{postBetaFit294MP7} Using mobility data M1$\sim$M6 and policy data H2, H3, H6, H7, train 294 days from Apr 4, 2020 to Jan 22, 2021; test 35 days from Jan 23, 2020 to Feb 26, 2021.}
\end{figure}

\begin{figure}[ht]
\centering
\includegraphics[width=0.6\textwidth]{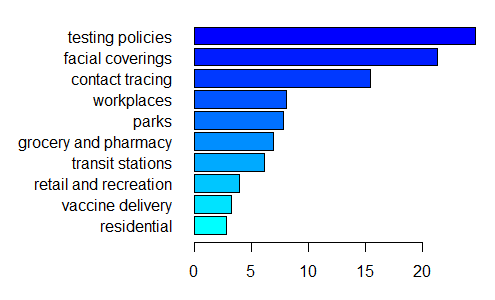}
\caption{\label{PostRI294MP7} Relative influence of mobility variables M1$\sim$M6 and policy variables H2, H3, H6, H7 when trained for 294 days from Apr 4, 2020 to Jan 22, 2020.}
\end{figure}

\begin{figure}[ht]
\centering
\includegraphics[width=0.49\textwidth]{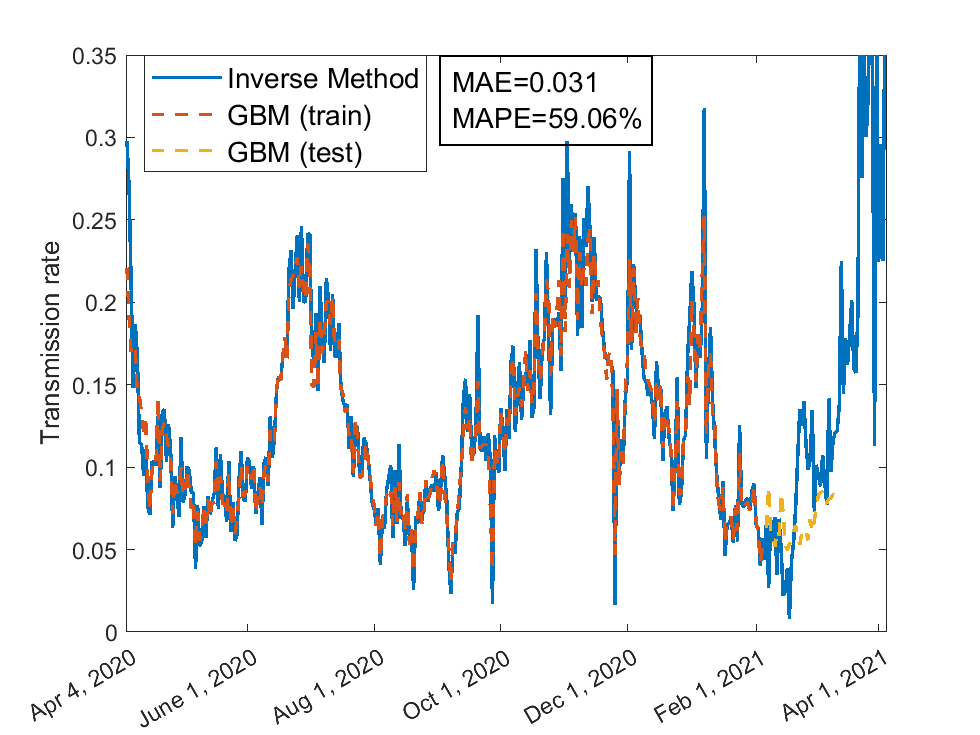}
\includegraphics[width=0.49\textwidth]{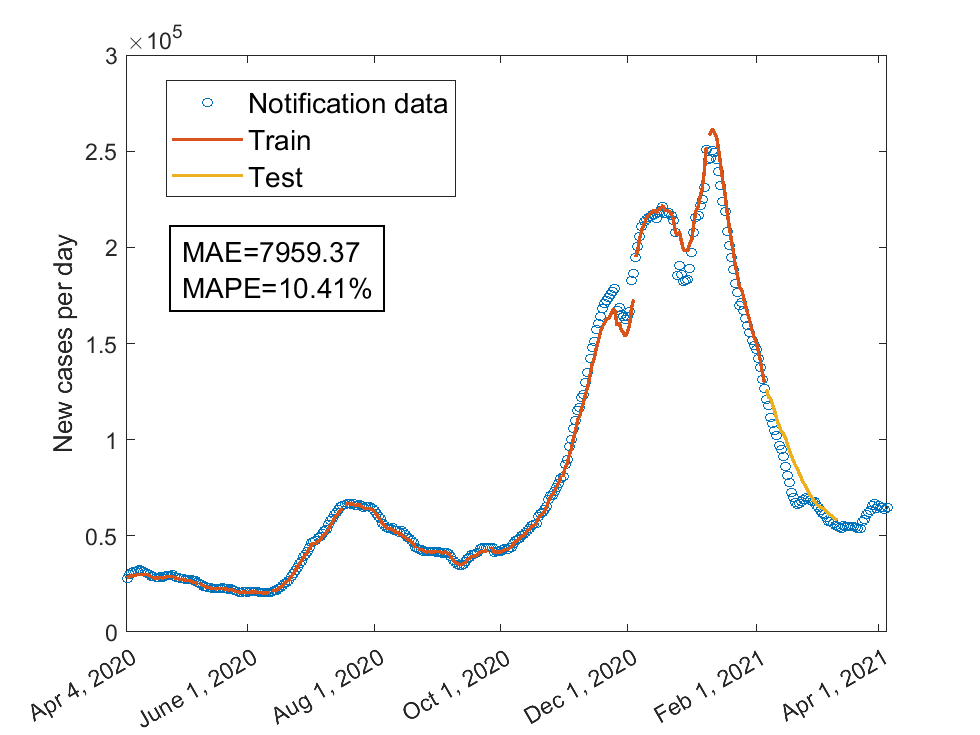}
\caption{\label{postBetaFit308MP7} Using mobility data M1$\sim$M6 and policy data H2, H3, H6, H7, train 308 days from Apr 4, 2020 to Feb 5, 2021; test 35 days from Feb 6, 2021 to Mar 12, 2021.}
\end{figure}

\begin{figure}[ht]
\centering
\includegraphics[width=0.6\textwidth]{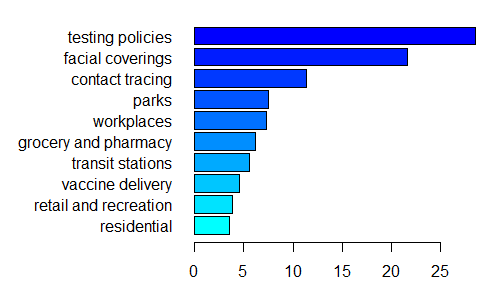}
\caption{\label{PostRI308MP7} Relative influence of mobility variables M1$\sim$M6 and policy variables H2, H3, H6, H7 when trained for 308 days from Apr 4, 2020 to Feb 5, 2021.}
\end{figure}

\begin{figure}[ht]
\centering
\includegraphics[width=0.49\textwidth]{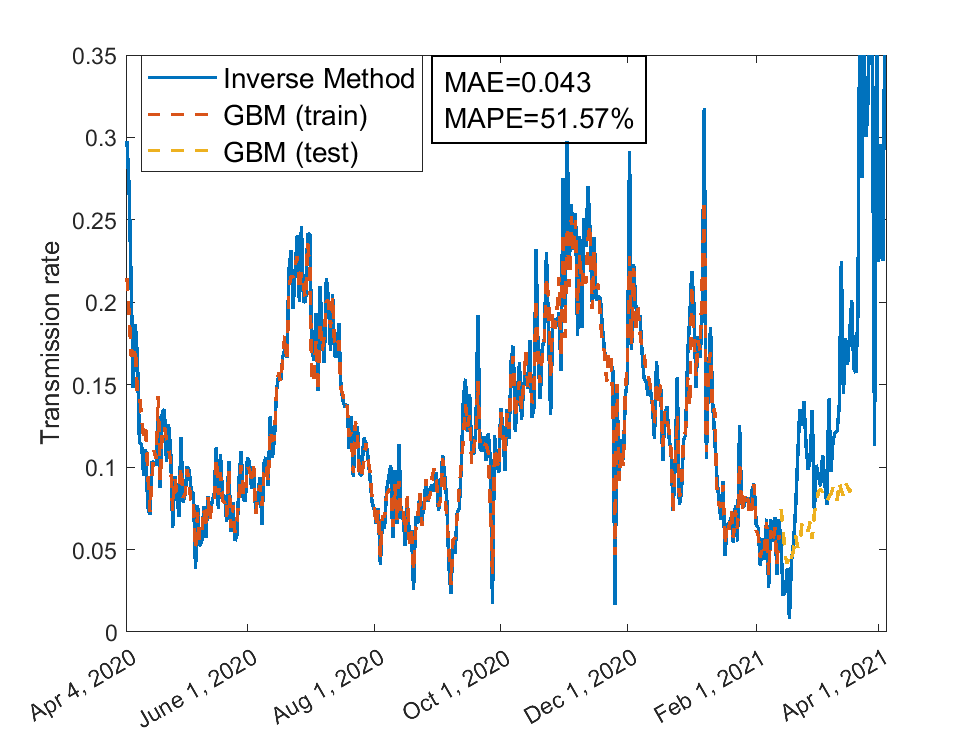}
\includegraphics[width=0.49\textwidth]{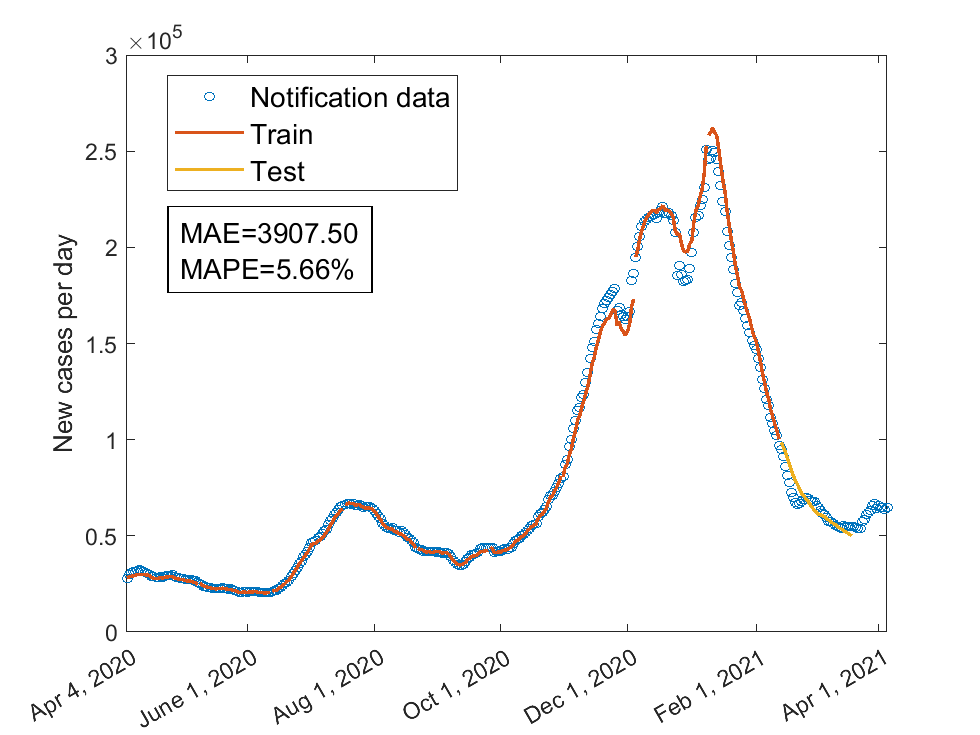}
\caption{\label{postBetaFit315MP7} Using mobility data M1$\sim$M6 and policy data H2, H3, H6, H7, train 315 days from Apr 4, 2020 to Feb 12, 2021; test 35 days from Feb 13, 2021 to Mar 19, 2021.}
\end{figure}

\begin{figure}[ht]
\centering
\includegraphics[width=0.6\textwidth]{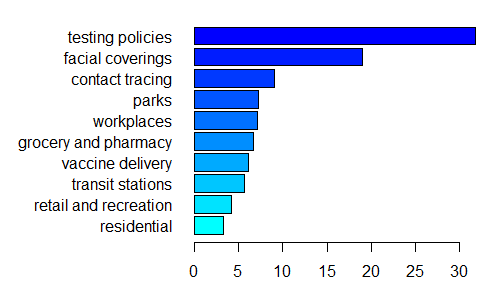}
\caption{\label{PostRI315MP7} Relative influence of mobility variables M1$\sim$M6 and policy variables H2, H3, H6, H7 when trained for 315 days from Apr 4, 2020 to Feb 12, 2021.}
\end{figure}

In this Appendix, we present supplementary figures and tables. 
The selected training and testing results about the transmission rates as well as the fittings with notification data of daily confirmed cases corresponding to the training durations are presented in Figures \ref{postBetaFit252P}, \ref{postBetaFit259P}, \ref{postBetaFit308P} for the model with policy as the only predictors, in Figures 
\ref{postBetaFit245M}, \ref{postBetaFit273M} for the model with mobility as the only predictors, and in Figures \ref{postBetaFit245MP7}, 
\ref{postBetaFit252MP7}, \ref{postBetaFit294MP7},  \ref{postBetaFit308MP7}, \ref{postBetaFit315MP7} for the model with both mobility and policy as predictors.
Each of these figures is followed by a table and a figure of the relative influence of the involved predictor variables in training the model.
Tables \ref{tablePost252P}, \ref{tablePost259P}, \ref{tablePost308P} and Figures \ref{PostRI252P}, \ref{PostRI259P}, \ref{PostRI308P} show the relative influence of the policy vairables when the model is trained for 252 days, 259 days, 308 days, respectively.
Tables \ref{tablePost245M}, \ref{tablePost273M} and  Figures \ref{PostRI245M}, \ref{PostRI273M}
present the relative influence of the mobility variables when the model is trained for 245 days, 273 days, respectively.
The relative influence of the mobility variables and part of the predictor variables are given in
Tables \ref{tablePost245MP7}, \ref{tablePost252MP7}, \ref{tablePost294MP7}, \ref{tablePost308MP7}, 
\ref{tablePost315MP7} and Figures
\ref{PostRI245MP7}, \ref{PostRI252MP7}, 
\ref{PostRI294MP7}, 
\ref{PostRI308MP7}, \ref{PostRI315MP7} when the model is trained for 245 days, 252 days, 294 days, 308 days, 315 days, respectively.

\begin{table}[ht]
\centering
\begin{tabular}{l|l}
Variable & Relative influence ($\%$)\\\hline
restrictions on gatherings & $33.909130$\\
workplace closing &        $11.232488$\\
school closing &           $11.028680$\\
stay at home requirements & $6.639884$\\
testing policies &         $6.416839$\\
protection of elderly  people & $6.063271$\\
contact tracing &          $5.313348$\\
cancel public events &     $5.160384$\\
facial coverings &         $4.374857$\\
close public transport &   $3.710009$\\
restrictions on internal movement & $3.174205$\\
public information campaigns  &     $2.976905$\\
international travel controls & $0.000000$\\
vaccination delivery &     $0.000000$
\end{tabular}
\caption{\label{tablePost252P} Relative influence of policy variables C1$\sim$C8, H1$\sim$H3, H6$\sim$H8 when trained for 252 days from Apr 4, 2020 to Dec 11, 2020.}
\end{table}

\begin{table}[ht]
\centering
\begin{tabular}{l|l}
Variable & Relative influence ($\%$)\\\hline
restrictions on gatherings & $34.39$\\
school closing &           $11.035040$\\
workplace closing &        $6.938807$\\
stay at home requirements & $6.589626$\\
cancel public events &     $6.576977$\\
testing policies &         $6.488583$\\
protection of elderly people & $6.356869$\\
contact tracing &          $6.088365$\\
public information campaigns &      $4.626977$\\
close public transport &   $4.137645$\\
facial coverings &         $3.805337$\\
restrictions on internal movement & $2.962269$\\
international travel controls & $0.000000$\\
vaccination delivery &     $0.000000$
\end{tabular}
\caption{\label{tablePost259P} Relative influence of policy variables C1$\sim$C8, H1$\sim$H3, H6$\sim$H8 when trained for 259 days from Apr 4, 2020 to Dec 18, 2020.}
\end{table}

\begin{table}[ht]
\centering
\begin{tabular}{l|l}
Variable & Relative influence ($\%$)\\\hline
restrictions on gatherings &        $30.7225696$\\
testing policies           &       $15.9295981$\\
facial coverings           &        $9.5674599$\\
school closing             &        $7.6215169$\\
stay at home requirements  &        $6.2229438$\\
protection of elderly people &      $5.8110858$\\
cancel public events       &        $5.4201751$\\
workplace closing          &        $5.3760962$\\
close public transport     &        $3.9418682$\\
contact tracing            &        $3.6403693$\\
vaccination delivery       &        $2.4611543$\\
restrictions on internal movement & $2.3506657$\\
public information campaigns &      $0.9344972$\\
international travel controls  &    $0.0000000$
\end{tabular}
\caption{\label{tablePost308P} Relative influence of policy variables C1$\sim$C8, H1$\sim$H3, H6$\sim$H8 when trained for 308 days from Apr 4, 2020 to Feb 5, 2021.}
\end{table}

\begin{table}[ht]
\centering
\begin{tabular}{l|l}
Variable & Relative influence ($\%$)\\\hline
parks & $32.023163$\\
workplaces & $19.337368$\\
transit stations & $15.651014$\\
grocery and pharmacy & $13.396451$\\
retail and recreation & $12.160510$\\
residential & $7.431494$
\end{tabular}
\caption{\label{tablePost245M} Relative influence of mobility variables M1$\sim$M6 when trained for 245 days from Apr 4, 2020 to Dec 4, 2020.}
\end{table}

\begin{table}[ht]
\centering
\begin{tabular}{l|l}
Variable & Relative influence ($\%$)\\\hline
parks & $30.198832$\\
workplaces & $18.280085$\\
transit stations & $14.981685$\\
retail and recreation & $14.348039$\\
grocery and pharmacy & $13.616837$\\
residential & $8.574522$
\end{tabular}
\caption{\label{tablePost273M} Relative influence of mobility variables M1$\sim$M6 when trained for 273 days from Apr 4, 2020 to Jan 1, 2021.}
\end{table}

\begin{table}[ht]
\centering
\begin{tabular}{l|l}
Variable & Relative influence ($\%$)\\\hline
testing policies & $32.095679$\\         facial coverings & $20.416952$\\          contact tracing & $12.275297$\\              workplaces & $8.506240$\\
parks & $8.419345$\\       transit stations & $8.237208$\\ 
retail and recreation & $4.503228$\\  
grocery and pharmacy & $3.228911$\\               residential & $2.317140$\\
vaccine delivery & $0.000000$
\end{tabular}
\caption{\label{tablePost245MP7} Relative influence of mobility variables M1$\sim$M6 and policy variables H2, H3, H6, H7 when trained for 245 days from Apr 4, 2020 to Dec 4, 2020.}
\end{table}

\begin{table}[ht]
\centering
\begin{tabular}{l|l}
Variable & Relative influence ($\%$)\\\hline
testing policies & $31.721798$\\
facial coverings & $21.118352$\\
contact tracing & $12.370792$\\      workplaces & $8.812522$\\
transit stations & $8.070839$\\
parks & $7.886678$\\
retail and recreation & $4.246667$\\
grocery and pharmacy & $3.402456$\\
residential & $2.369897$\\
vaccine delivery & $0.000000$
\end{tabular}
\caption{\label{tablePost252MP7} Relative influence of mobility variables M1$\sim$M6 and policy variables H2, H3, H6, H7 when trained for 252 days from Apr 4, 2020 to Dec 11, 2020.}
\end{table}

\begin{table}[ht]
\centering
\begin{tabular}{l|l}
Variable & Relative influence ($\%$)\\\hline
testing policies & $24.657548$\\
facial coverings & $21.269020$\\
contact tracing & $15.376631$\\
workplaces & $8.073849$\\
parks & $7.799143$\\
grocery and pharmacy & $6.908603$\\
transit stations & $6.076437$\\
retail and recreation & $3.897032$\\
vaccine delivery & $3.183959$\\
residential & $2.757777$
\end{tabular}
\caption{\label{tablePost294MP7} Relative influence of mobility variables M1$\sim$M6 and policy variables H2, H3, H6, H7 when trained for 294 days from Apr 4, 2020 to Jan 22, 2020.}
\end{table}

\begin{table}[ht]
\centering
\begin{tabular}{l|l}
Variable & Relative influence ($\%$)\\\hline
testing policies & $28.575819$\\
facial coverings & $21.600347$\\
contact tracing & $11.359680$\\       parks & $7.532694$\\
workplaces & $7.257706$\\
grocery and pharmacy & $6.156383$\\
transit stations & $5.582075$\\
vaccine delivery & $4.550569$\\
retail and recreation & $3.825492$\\
residential & $3.559234$
\end{tabular}
\caption{\label{tablePost308MP7} Relative influence of mobility variables M1$\sim$M6 and policy variables H2, H3, H6, H7 when trained for 308 days from Apr 4, 2020 to Feb 5, 2021.}
\end{table}

\begin{table}[ht]
\centering
\begin{tabular}{l|l}
Variable & Relative influence ($\%$)\\\hline
testing policies & $31.865308$\\
facial coverings & $18.987057$\\
contact tracing & $9.021340$\\
parks & $7.267074$\\
workplaces & $7.127695$\\
grocery and pharmacy & $6.679093$\\
vaccine delivery & $6.087163$\\
transit stations & $5.594120$\\
retail and recreation & $4.116551$\\
residential & $3.254600$
\end{tabular}
\caption{\label{tablePost315MP7} Relative influence of mobility variables M1$\sim$M6 and policy variables H2, H3, H6, H7 when trained for 315 days from Apr 4, 2020 to Feb 12, 2021.}
\end{table}

\end{document}